\documentclass[sigconf, nonacm]{acmart}

\usepackage{listings}
\usepackage{scalefnt}
\usepackage{tikz}
\usepackage{xcolor}
\usepackage{subfigure}
\usepackage{xspace}
\usepackage{multirow}
\usepackage{enumitem}
\definecolor{mygreen}{rgb}{0,0.6,0}
\definecolor{mygray}{rgb}{0.5,0.5,0.5}
\definecolor{mymauve}{rgb}{0.58,0,0.82}
\makeatletter
\def\@ACM@checkaffil{% Only warnings <<<<<<<<<<<<<<<<
    \if@ACM@instpresent\else
    \ClassWarningNoLine{\@classname}{No institution present for an affiliation}%
    \fi
    \if@ACM@citypresent\else
    \ClassWarningNoLine{\@classname}{No city present for an affiliation}%
    \fi
    \if@ACM@countrypresent\else
        \ClassWarningNoLine{\@classname}{No country present for an affiliation}%
    \fi
}
\makeatother

\AtBeginDocument{%
  }

\setcopyright{acmcopyright}
\copyrightyear{2023}
\acmYear{2023}
\acmDOI{XXXXXXX.XXXXXXX}

\acmConference[Conference SIGMOD 2024]{Make sure to enter the correct
  conference title from your rights confirmation emai}{June 09--15,
  2024}{Santiago, Chile}
\acmPrice{15.00}
\acmISBN{978-1-4503-XXXX-X/18/06}

\begin{document}
\pagestyle{plain} 
%%
%% The "title" command has an optional parameter,
%% allowing the author to define a "short title" to be used in page headers.
\title{GraphScope Flex: LEGO-like Graph Computing Stack}

%%
%% The "author" command and its associated commands are used to define
%% the authors and their affiliations.
%% Of note is the shared affiliation of the first two authors, and the
%% "authornote" and "authornotemark" commands
%% used to denote shared contribution to the research.

% \author{Lars Th{\o}rv{\"a}ld}
% \affiliation{%
%   \institution{The Th{\o}rv{\"a}ld Group}
%   \streetaddress{1 Th{\o}rv{\"a}ld Circle}
%   \city{Hekla}
%   \country{Iceland}}
% \email{larst@affiliation.org}

\author{Tao He, Shuxian Hu, Longbin Lai, Dongze Li, Neng Li, 
Xue Li, Lexiao Liu, Xiaojian Luo, Binqing Lyu, Ke Meng, 
Sijie Shen, Li Su, Lei Wang, Jingbo Xu, 
Wenyuan Yu, Weibin Zeng, Lei Zhang, Siyuan Zhang, 
Jingren Zhou, Xiaoli Zhou, Diwen Zhu
}
\affiliation{%
  \institution{Alibaba Group}
  % \streetaddress{30 Shuangqing Rd}
  % \city{Haidian Qu}
  % \state{Beijing Shi}
  % \country{China}
  }
\email{graphscope@alibaba-inc.com}

% \author{Charles Palmer}
% \affiliation{%
%   \institution{Palmer Research Laboratories}
%   % \streetaddress{8600 Datapoint Drive}
%   % \city{San Antonio}
%   % \state{Texas}
%   \country{USA}
%   % \postcode{78229}
%   }

%%
%% By default, the full list of authors will be used in the page
%% headers. Often, this list is too long, and will overlap
%% other information printed in the page headers. This command allows
%% the author to define a more concise list
%% of authors' names for this purpose.
\renewcommand{\shortauthors}{He et al.}

\newcommand{\stitle}[1]{\vspace{0.5ex}\noindent{\bf #1}}
\newcommand{\etitle}[1]{\vspace{0.5ex}\noindent{\em\underline{#1}}}
\newcommand{\eetitle}[1]{\vspace{0.5ex}\noindent{\em{#1}}}
\newcommand{\eat}[1]{}

\newcommand{\kw}[1]{\textsf{#1}}
\newcommand{\gsf}{{\kw{GraphScope Flex}}}
\newcommand{\gs}{{\kw{GraphScope}}}
\newcommand{\grape}{{\kw{GRAPE}}}
\newcommand{\flash}{{\kw{FLASH}}}
\newcommand{\grin}{{\kw{GRIN}}}
\newcommand{\gart}{{\kw{GART}}}
\newcommand{\ingress}{{\kw{Ingress}}}
\newcommand{\gaia}{{\kw{Gaia}}}
\newcommand{\glogue}{\kw{GLogue}}
\newcommand{\hiactor}{{\kw{HiActor}}}
\newcommand{\vyd}{{\kw{Vineyard}}}
\newcommand{\graphar}{{\kw{GraphAr}}}
\newcommand{\flexbuild}{{\kw{flexbuild}}}
\newcommand{\gl}{{\kw{GraphLearn}}}

\newcommand{\code}[1]{\texttt{#1}}
\newcommand{\traits}{{\code{traits}}}
\newcommand{\trait}{{\code{trait}}}

% suggest to use code style for the operators.
\newcommand{\kk}[1]{\texttt{#1}}
\newcommand{\expandedge}{{\kk{EXPAND}\_\kk{EDGE}}}
\newcommand{\getvertex}{{\kk{GET}\_\kk{VERTEX}}}
\newcommand{\matchstart}{{\kk{MATCH}\_\kk{START}}}
\newcommand{\matchend}{{\kk{MATCH}\_\kk{END}}}
\newcommand{\pathstart}{{\kk{PATH}\_\kk{START}}}
\newcommand{\pathend}{{\kk{PATH}\_\kk{END}}}
\newcommand{\project}{\kk{PROJECT}}
\newcommand{\select}{\kk{SELECT}}
\newcommand{\order}{\kk{ORDER}}
\newcommand{\group}{\kk{GROUP}}
\newcommand{\map}{\kk{MAP}}
\newcommand{\flatmap}{\kk{FLATMAP}}
\newcommand{\source}{\kk{SOURCE}}

\newcommand{\ie}{\emph{i.e.,}\xspace}
\newcommand{\eg}{\emph{e.g.,}\xspace}
\newcommand{\wrt}{\emph{w.r.t.}\xspace}
\newcommand{\aka}{\emph{a.k.a.}\xspace}
\newcommand{\kwlog}{\emph{w.l.o.g.}\xspace}
\newcommand{\kwhp}{\emph{w.h.p.}\xspace}

\newcommand{\todo}[1]{\textcolor{red}{$\Rightarrow$#1}}
\newcommand{\tbf}{\textbf{\textcolor{red}{X}}\xspace}
\newcommand{\warn}[1]{{\color{red}{#1}}}
\newcommand{\revise}[1]{{\textcolor{black}{#1}}}

\newcommand*\circled[1]{\tikz[baseline=(char.base)]{
                \node[shape=circle,draw,inner sep=0.6pt] (char) {#1};}}

\newcommand*\circledd[1]{\tikz[baseline=(char.base)]{
            \node[shape=circle,draw,inner sep=0.3pt] (char) {\scalefont{0.9}{#1}};}}
%%
%% The abstract is a short summary of the work to be presented in the
%% article.
\eat{\begin{abstract}

Graph computing is prevalent,
and numerous systems are proposed to process large-scale graph data.
Among these, \gs~ stands out as the first one-stop system
covering multiple types of graph computing,
including graph traversal, graph analytics,
and graph learning.
However, our experience with \gs~ has taught us that
a \emph{``one-size-fits-all''} approach is overly ambitious,
facing challenges due to diverse programming interfaces,
applications, and data storage formats.
In this paper, we introduce \gsf,
a LEGO-like graph computing stack designed to address these complexities.
\gsf~ adopts a modular and disaggregated design,
aiming to minimize both resource and cost requirements
while providing a flexible and user-friendly deployment experience.
We detail the architecture and key design aspects of \gsf,
and experimentally verify its effectiveness and efficiency
on synthetic data and in real-world applications.
\end{abstract}}

\begin{abstract}
  Graph computing has become increasingly crucial in processing
  large-scale graph data, with numerous systems developed for this purpose. Two years ago, we introduced \gs~ as a system addressing a wide array of graph computing needs, including graph traversal, analytics, and learning in one system. Since its inception,~\gs~ has achieved significant technological advancements and gained widespread adoption across various industries. However, one key lesson from this journey has been understanding the limitations of a \emph{``one-size-fits-all''} approach, especially when dealing with the diversity of programming interfaces, applications, and data storage formats in graph computing. In response to these challenges, we present \gsf, the next iteration of \gs. \gsf~is designed to be both resource-efficient and cost-effective, while also providing flexibility and user-friendliness through its LEGO-like modularity. This paper explores the architectural innovations and fundamental design principles of \gsf, all of which are direct outcomes of the lessons learned during our ongoing development process. We validate the adaptability and efficiency of \gsf~ with extensive evaluations on synthetic and real-world datasets. \revise{
    The results show that \gsf~ achieves $2.4\times$ throughput and up to $55.7\times$ speedup over other systems on the LDBC Social Network and Graphalytics benchmarks, respectively. Furthermore, \gsf~ accomplishes up to a 2,400$\times$ performance gain in real-world applications, demonstrating its proficiency across a wide range of graph computing scenarios with increased effectiveness.  
  }
  \eat{The demonstrating its proficiency in handling a broad spectrum of graph computing scenarios with enhanced effectiveness.}
  % \vspace*{-2ex}
\end{abstract}

%%
%% The code below is generated by the tool at http://dl.acm.org/ccs.cfm.
%% Please copy and paste the code instead of the example below.
%%
\eat{

\begin{CCSXML}
<ccs2012>
   <concept>
       <concept_id>10010520.10010521.10010537.10010539</concept_id>
       <concept_desc>Computer systems organization~n-tier architectures</concept_desc>
       <concept_significance>300</concept_significance>
       </concept>
 </ccs2012>
\end{CCSXML}

\ccsdesc[300]{Computer systems organization~n-tier architectures}

%%
%% Keywords. The author(s) should pick words that accurately describe
%% the work being presented. Separate the keywords with commas.
\keywords{graph computing, distributed system, graph analytics}
}
% \received{20 February 2007}
% \received[revised]{12 March 2009}
% \received[accepted]{5 June 2009}

%%
%% This command processes the author and affiliation and title
%% information and builds the first part of the formatted document.
\maketitle

\section{Introduction}
\label{sec:intro}

Graphs are increasingly becoming the backbone of numerous real-world applications, 
permeating diverse fields such as social networks, e-commerce, bioinformatics, fintech, 
and knowledge base~\cite{hierarchical-knowledge-graph, zamini2022review, azadifar2022graph, li2022graph, tu2023disentangled}. These applications often involve complex interactions and relationships, 
uniquely represented through graph structures. As the scale of graph data and the intricacy of applications 
grow, the demand for specialized graph processing systems has escalated. Systems like Pregel~\cite{malewicz2010pregel}, 
Spark GraphX~\cite{graphx}, TuGraph~\cite{tugraph}, TigerGraph~\cite{tigergraph} and PyG~\cite{pyg} have been developed to address specific graph computation requirements.
However, developers often find themselves juggling 
multiple systems with vastly different programming models and runtime environments. This multiplicity gives 
rise to a host of issues, such as managing the complexities of data representation, resource scheduling, 
and performance tuning across these disparate systems.

In response to these challenges, \gs~\cite{fan2021graphscope} was developed as a pioneering solution, 
designed to offer a comprehensive approach to graph computing. Unlike its predecessors, \gs~supports a wide 
range of computing paradigms, enabling it to handle diverse tasks like graph analytics, graph learning, 
and interactive queries within a single, unified system. Internally, \gs~employs an extension of Gremlin 
and a unified dataflow engine, capable of processing various types of graph computations efficiently. 
Furthermore, it enhances interoperability with other frameworks in the PyData ecosystem, offering an 
integrated solution that reduces the burden of working with multiple graph computing systems.
Despite its ambition to be an efficient, user-friendly, and all-encompassing solution for graph computing, 
\gs~has encountered practical challenges.
% These challenges stem from the diversity and complexity inherent 
%in real-world graph computing requirements, where a ``one-size-fits-all'' approach often falls short in 
%addressing the nuanced demands of different applications.

\begin{figure}[t!]
    \centering
    % \vspace*{-1.5em}
    \includegraphics[width=\linewidth]{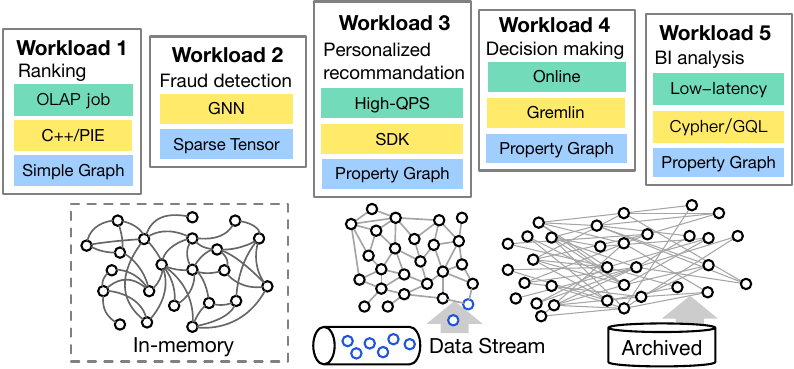}
    \vspace*{-2em}
    \caption{Diversified graph workloads in e-commerce.}
    \label{fig:workloads}
    \vspace{-2ex}
\end{figure}

\eat{%eat}
Graphs are ubiquitous in many real-world applications, such as social networks, e-commerce, bioinformatics, fintech, and knowledge graphs. 
With the increasing size of graphs and the complexity of applications, a variety of specialized graph processing systems have been developed. 
These include Pregel~\cite{malewicz2010pregel}, Spark GraphX, TuGraph, TigerGraph, and PyG, each designed to tackle specific types of graph computations. 
Unlike previous systems, \gs~\cite{fan2021graphscope} is a pioneering and comprehensive system that supports multiple computing paradigms. It is capable of handling graph analytics, graph learning, and interactive queries within a single system. 
Internally, \gs~employs an extension of Gremlin and a unified dataflow engine to process various types of graph computations. Furthermore, it offers interoperability with other frameworks in the PyData ecosystem. 
However, despite its aim to be an efficient, user-friendly, and one-stop solution, \gs~encounters its own challenges in practice.
}%eat

%Figure~\ref{fig:workloads} illustrates a simplified real-world example of 
%diversified graph computing workloads in Alibaba. 
%Data from various sources are modeled as graphs, with vertices representing buyers, sellers, items or other entities. Edges denote relationships, including buying, selling, and reviewing. These graphs can be dynamic or static, large or small, and may fit in memory or serve as large archives of historical data.

\stitle{Real-world example.}
Figure~\ref{fig:workloads} offers a simplified portrayal of the diversity inherent in graph computing workloads as seen in real-world settings with Alibaba as an example. Here, data is represented as graphs. Within these graphs, vertices symbolize entities such as buyers, sellers, and items. In contrast, edges represent relationships or activities such as buying, selling, or reviewing. These graphs could be either dynamic or static, large or small, and either fit in-memory or exist as extensive historical archives.

These diverse graphs inhabit different graph workloads tuned to cater to a variety of business scenarios. For instance, a ranking workload stipulates the order in which items or sellers appear in search results (marked as Workload 1 in Figure~\ref{fig:workloads}). On the other hand, to unravel suspicious entities potentially participating in fraudulent activities to manipulate their ratings and rankings, an anti-fraud task deploying link prediction (SEAL) algorithm is activated (Workload 2). The platform also offers a personalized recommendation service predicting user interests based on historical behaviors and tastes (Workload 3), while a BI analysis-enabled WebUI equips data scientists with the tools necessary to make data-driven decisions (Workload 4), such as identifying the most popular sellers or items in specific regions. Furthermore, some ad-hoc BI workloads sporadically operate over large historical graph data archives (Workload 5), which often prove too sizable to fit into the memory of a running graph database given their infrequent usage and high costs.

\eat{
Various workloads were conducted on different graphs to cater to different business scenarios. 
For instance, the order in which items or sellers appear in search results is given by a ranking workload (Shown as Workload 2 in Figure~\ref{fig:workloads}).
A minority of sellers and buyers might collaborate in fraudulent transactions and reviews to artificially boost their ratings and rankings. 
To detect these suspicious entities, an anti-fraud task with data mining techniques might be employed (Workload 1).
Furthermore, the platform offers a personalized recommendation service, which predicts user interests based on their historical behavior and their friends' preferences (Workload 3). Additionally, there is a WebUI for BI analysis, enabling data scientists to explore the graph and make informed decisions (Workload 4), such as identifying the most popular items or sellers in a specific region. In some cases, these ad-hoc BI workloads are occasionally to be conducted over a large historical archival of graph data (Workload 5), which is too costly to fit in memory of a running graph database for such low frequency needs.}

This example -- to some extent mirroring the design motif behind \gs~-- further reasserts the reality that real-world graph computing workloads are diverse, yet they share certain characteristics. Workload 2 encapsulates these shared traits via its utilization of an algorithm akin to PageRank— a category of graph analytics usually demanding high data-intensity and being memory-bound. On the other hand, Workload 1 involves a Graph Neural Network (GNN) task, which often necessitates memory-bound data sampling and CPU/GPU resource-limited computation-intensive back-end training. In contrast, Workloads 4 and 5 are frequently interactive complex query-based graph exploration tasks which typically target a smaller selective subset of the graph based on certain conditions.

Even though \gs~offers support for distributed immutable in-memory graph storage suitable for static graph processing, real-life workloads often demand varying formats and access paths to the data. As illustrated in Figure~\ref{fig:workloads}, visible in the blue section, iterative analytics like k-core and label propagation generally operate over attributed or simple graphs. Conversely, GNN models work on sparse tensors, while interactive queries use a labeled property graph model to facilitate complex attribute-oriented queries. Furthermore, data lifecycle stages also differ — data mining and GNN training usually occur on a snapshot of graph data, while GNN inference and interactive queries function on dynamic graphs receiving continuous updates. BI analysis can occur either over dynamic graphs or historical graph data archives.

While the unified interface proposition from \gs~might seem beneficial, in practice its one-size-fits-all strategy is ineffective due to the disparate needs of various workloads and user preferences. For instance, data scientists performing interactive queries often prefer domain-specific languages such as Gremlin or Cypher, while developers working on GNN models lean towards specialty libraries such as PyG. Similarly, services necessitating high query throughput might use parameterized queries structured as stored procedures and for graph analytics, iterative algorithms are typically coded in C++ or Java and made accessible as built-in libraries.

Our experience with \gs~has underscored the limitations of a ``one-size-fits-all'' approach in the complex arena of graph computing. While requirements can \emph{range widely}, they often \emph{exhibit underlying commonalities}. We've found that a design choice effective for one scenario may fall short in another. No single solution — whether concerning the engine, interface, or storage — can accommodate all varied requirements comprehensively. This divergence between theory and practice revealed through our work on \gs, which aimed for a unified solution, led to frustrating trade-offs and imposed certain shortcomings onto specific tasks.

\stitle{\gsf.} In this paper, we introduce ~\gsf, the next iteration of \gs. Adopting a modular architecture, ~\gsf~ aims to minimize resource and cost overhead while enhancing deployment flexibility and user experience. Unlike \gs's unified dataflow engine, ~\gsf~ disaggregates the engines for various graph tasks, as well as their interfaces and storages. This modular design makes the system architecture akin to a set of LEGO bricks: users can selectively deploy components of ~\gsf~ to streamline the deployment process and adapt the system to their specific needs. 

In summary, we make the following contributions:

\begin{itemize}[noitemsep,topsep=0pt,leftmargin=*]
\itemsep0em
    \item We provide a comprehensive exploration of the diverse graph models and storage schemes in graph computing, the different graph computing workloads and their applications, an array of programming interfaces used across graph computing, the variety of performance requirements and existing graph computing systems. And we discuss the opportunities to decouple common components and makes them composable to meet the diverse needs of graph computing (\S 2).
    \item We offer a bird-eye view of the architecture of \gsf~ and an in-depth discussion on \revise{its modules and techniques.(\S 3)}. \eat{the strategies and techniques employed by \gsf (\S 3).}
    \item We delve into \grin~ for decoupling graph storage from workloads, and key graph storage techniques, such as in-memory \vyd, \revise{dynamic graph store \gart~}and archive format \graphar (\S 4). 
    \item We demonstrate the flexibility that \gsf~ offers in handling graph querying workloads. \revise{~\gsf~ uses an intermediate representation to support both Gremlin and Cypher, translating and executing queries on appropriate computing engines based on their OLAP or OLTP characteristics (\S 5).} 
    \item We scrutinize the manner in which \gsf~ favors the handling of graph analytics workloads. \revise{It provides a variety of built-in algorithms and programming interfaces, complemented by support for both CPU and GPU backends (\S 6)}.
    \item We examine the suitability of \gsf~ in catering to diverse requirements in GNN workloads, \revise{highlighting its decoupling of sampling and training for independent scaling (\S 7).} 
    \item \eat{We present case studies demonstrating the practical application of \gsf~ and its efficacy and versatility in real-world scenarios (\S 8).\tbf} \revise{We present case studies illustrating \gsf's practical application, efficacy, and versatility in real-world scenarios (\S 8).}
    \eat{\item Lastly, we present empirical data to evaluate the performance, flexibility, and ease of use of \gsf, showing \gsf achieves $2.4\times~$ throughput and up to $55.7~\times~$ speedup over the state-of-the-art systems on LDBC Social Network Benchmark and Graphalytics benchmark, respectively. Better still, it performs well in real-world applications, achieves up to $2,400\times~$ speedup in performance. (\S 9).}
    \item Lastly, we provide empirical data assessing \gsf's performance and flexibility. \revise{Our findings show that \gsf~ attains $2.4\times$ throughput and up to $55.7\times$ speedup compared to the state-of-the-art systems on the LDBC Social Network and Graphalytics benchmarks, respectively. Moreover, \gsf~ excels in real-world applications, achieving up to 2,400$\times$ performance speedup compared to previous solutions (\S 9)}.
\end{itemize}

\eat{
\begin{itemize}
\item We introduce a comprehensive graph computing system designed to meet the demands of interactive querying, graph analytics, graph learning, and pattern matching. Leveraging a LEGO-like modular architecture, our system offers the flexibility and cost-effectiveness essential for handling complex graph computing workflows in real-world applications. 
\item In \gsf, we propose a unified parser and optimizer for graph traversal workloads. The parser can accept queries in either Gremlin or Cypher and produces a unified Intermediate Representation (IR). Subsequently, the optimizer generates an optimized execution plan tailored for appropriate backends. These backends include a data-parallel engine optimized for handling long and complex queries with low latency, as well as a high-performance actor-based engine designed for high-query-per-second (QPS) use-cases.
\item In \gsf, we introduce a unified storage interface that allows users to select an appropriate storage backend tailored to their specific needs, while maintaining transparency at the engine layer for seamless operation. 
\item Additionally, we offer a graph format to facilitate data exchange between \gsf~ and external systems. This enables seamless construction and conversion of graph data layouts, thereby extending \gsf's interoperability within broader big data ecosystems. 
\item Lastly, we present a variety of typical use cases for \gsf~ and experimentally verify that \gsf~ is not only straightforward to deploy, but also exhibits high performance in multiple real-world applications. 
\end{itemize}
}%eat
% paper organization:

% system architecture, challenges
% C1: m2m storage
% C2: high qps and high throughput
% C3: workflow with other systems in big data

% Design and Implementations
% D1: GRIN
% D2: unified optimizer
% D3: Python Interface and GraphAr

% User cases

% performances
% E1: GRIN
% E2: LDBC workload interactive
% E3: GraphAr
% E4: other use cases 
% related work & conclusion

\section{Background \& Related Work}
\label{sec-background}

In this section, we present the background on graph data models,
storage formats, applications, 
programming interfaces, performance requirements, and existing graph computing systems.
We also review the diversity within these components and 
highlight opportunities for a disaggregated system design.

\subsection{Graph Models \& Organizations}
Graphs are ubiquitous in real life, effectively modeling complex systems and relationships through vertices and edges. However, the superficial simplicity of graph abstraction conceals a landscape filled with diversity and fragmented nuances. Upon closer inspection, the concept of a graph becomes more complex and diverse than the relational data processing landscape.

Graphs can adopt various data models such as simple graphs, weighted graphs common in graph analytics tasks (like PageRank or SSSP)~\cite{zhang2006introduction,graphx,malewicz2010pregel,giraph}, 
sparse matrix/tensor used in GNN models~\cite{zhou2020graph,tf,pyg}, 
RDF observed in knowledge bases~\cite{decker2000semantic, rdf,ontotext}, 
or Labeled Property Graph (LPG)~\cite{angles2023pg, anikin2019labeled,baken2020linked} 
widely used in graph databases~\cite{tigergraph, tugraph, neo4j,janusgraph}. 
\eat{Each of these models supports different operations and access interfaces.} %eat: repeated in the next paragraph
Some common graph models are:

\begin{itemize}[noitemsep,topsep=0pt,leftmargin=*]
\itemsep0em
    \item \textit{Simple Graphs}: This basic model consists of vertices and edges, with each edge links two vertices. This model is unweighted and does not allow for additional metadata or properties. And the interfaces are generally of explorations along vertices and edges.
    \item \textit{Weighted Graphs}: These graphs enhance the simple graph model by assigning weights or costs to each edge. The weights can symbolize various metrics, from distance to connection strength.
    \item \textit{Sparse Matrix/Tensor}: Devised for Graph Neural Network (GNN) models, this type of graph considers vertices as matrix rows and edges as non-zero entries in the matrix. This method is generally adopted for high-dimensional data representation, and the interfaces are of matrix/tensor operations.
    \item \textit{Resource Description Framework (RDF)}: RDFs consist of triples - \emph{subject, predicate}, and \emph{object} -- that denote relationships (\emph{predicates}) between two entities or nodes (\emph{subject} and \emph{object}). This model facilitates the representation and integration of data from disparate sources, commonly used in knowledge bases.
    \item \textit{Labeled Property Graph (LPG)}: Widely embraced in graph databases, LPGs include labels and properties (key-value pairs) in the vertices and edges. Labels categorize vertices and edges, and properties furnish extra information, enhancing querying capabilities.
\end{itemize}

Each of these models supports varying operations and provides different access interfaces. Correspondingly, the storage and processing needs also differ substantially, thereby highlighting the need for customizable and flexible solutions in graph computing.

Aside from data models, the methods of storing and organizing graphs also diverge depending on specific requirements. For instance, graph analytics often call for iterative random accesses across the entire graph, making such tasks memory-bound. In these cases, storing the graph in-memory in a cache-friendly manner enhances performance. Graph analytics applications often
have no requirement for transactional updates and are performed on graphs that are periodically updated in batches~\cite{macko2015llama, terrace, dhulipala2019low}. While, some graph databases necessitate frequent updates and must maintain Atomicity, Consistency, Isolation, Durability (ACID) properties, demanding a suitable storage method~\cite{zhu13livegraph, Sortledton, de2021teseo, ediger2012stinger, kumar2020graphone}. At other times, the graph may be of archive nature -- extremely large and with low access frequency. Certain scenarios may permit single-machine operations, but often, graphs exceed the capacity of a single machine, requiring partitioned and distributed processing.

The plurality of graph data models when combined with the myriad ways of graph organization amplifies the diversity, making the landscape of graph processing far from \eat{homogenous or }standardized.

\begin{figure}[t!]
    \centering
    \includegraphics[width=\linewidth]{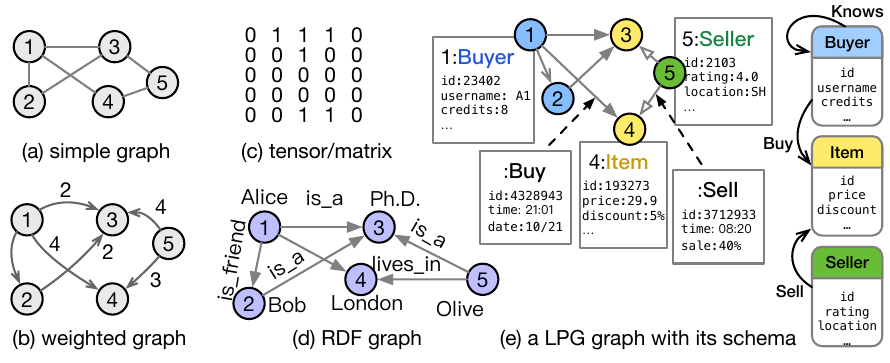}
    \vspace*{-2em}
    \caption{Graph models.}
    \label{fig:graphs}
    \vspace{-2em}
\end{figure}

\eat{

\stitle{Homogeneous and heterogeneous graph.} 
A homogeneous graph contains only one type of vertex and edge, as shown in Figure~\ref{fig:graphs}~(a).
Such graphs are often adequate for a wide range of traditional iterative algorithms 
that primarily focus on graph topology structure.
Most common graph algorithms we learned from textbooks 
such as shortest path, PageRank, centrality measures and community detection 
can operate efficiently on homogeneous graphs.
However, for certain GNN models and graph learning tasks, 
heterogeneous graphs as shown in Figure~\ref{fig:graphs}~(b), 
which include multiple types of vertices or edges, are more suitable. 
These heterogeneous graphs preserve more information, 
allowing for richer feature extraction during sampling for model training or inference.

\stitle{Property graph.} 
A property graph is a multi-typed graph wherein 
each vertex and edge is associated with a set of properties in key-value pairs
Similar to relational data models, property graphs adhere to a predefined schema. 
Figure~\ref{fig:graphs}~(c) and (d) depict an example property graph alongside its schema. 
This particular graph comprises three types of vertices—\textit{Buyer}, \textit{Item}, and \textit{Seller}—and 
two types of edges, \textit{Buy} and \textit{Sell}. 
Each type of vertex and edge carries a unique set of properties; 
for instance, \textit{Item} vertices possess properties such as \textit{id}, \textit{price}, and \textit{discount}. 
Property graphs are prevalent in graph databases 
due to their ability to encapsulate rich properties in graph.

In addition to the graph model, 
other factors such as access patterns and storage medium 
also need to be taken into consideration for the design of the storage abstraction.

\stitle{Access pattern.}
Given the diverse range of access patterns for graph data, 
a multitude of storage design choices are necessitated. 
For example, static and immutable graphs might best be served 
by a Compressed Sparse Row/Column (CSR/CSC) layout, 
which is optimal for high-speed random read access. 
In contrast, dynamic graphs particularly concern about the efficiency of updates. 
Furthermore, some applications, such as graph databases, require transactional support,
necessitating the storage adhere to ACID property.

\stitle{Medium.} 
The storage medium used for graph significantly influences the design 
of the storage abstraction layer. 
Graph data can be stored in various locations, 
ranging from in-memory storage and local disks to network-based solutions. 
In computing-intensive scenarios, 
storing graph data in GPU memory may be necessary to achieve ultra-high speeds. 
While in some other scenarios, 
applications may prioritize cost-effectiveness, 
opting for network-based storage services like S3 or OSS. 
It's important to note that data layouts vary between GPU and CPU, 
and the layout optimized for efficient network transfer may differ substantially. 
Furthermore, 
the data in medium should be structured in a way that facilitates easy sharing with other systems. 
Accordingly, 
storage design must take into account this diversity and be optimized for specific objectives.
}

\subsection{Applications of Graph Computing}

Graph computing covers a wide range of applications across various domains. 
We categorize these applications into three main types, each associated with specific graph models and querying paradigms.

\stitle{Graph Querying.}
Graph querying involves using specialized languages such as Gremlin~\cite{rodriguez2015gremlin}, Cypher~\cite{francis2018cypher}, GQL~\cite{gql}, and SPARQL~\cite{sparql} to interrogate and manipulate graph structures. This category primarily includes operations related to pattern matching and complex query formulations. Gremlin and Cypher are often employed for traversal and pattern matching in labeled property graphs, facilitating intricate queries and analyses. On the other hand, SPARQL is predominantly used with RDF graphs, focusing more on sophisticated pattern matching, data aggregation, and integration across various data sources. This distinction highlights the varied nature of graph querying, necessitating flexible and adaptable query processing capabilities in graph computing systems.

\stitle{Graph Analytics.}
\eat{Graph analytics focuses on analyzing the global structure and properties of graphs, using algorithms for clustering, centrality measures, shortest paths, and reachability. }
\revise{Graph analytics investigates global structure of graphs using algorithms for clustering, centrality, shortest paths, and reachability.} Applications in this category typically utilize simple or weighted graphs. For instance, clustering algorithms like Louvain~\cite{blondel2008fast} and centrality measures such as PageRank~\cite{page1998pagerank} are central to understanding network dynamics and influence patterns in domains like social network analysis and epidemiology ~\cite{holme2017three}.

\stitle{Graph Learning.}
Graph learning, especially through Graph Neural Networks (GNNs), applies machine learning techniques to graph-structured data, often represented as sparse tensors or matrices~\cite{zhou2020graph}. This process typically involves three key steps: sampling, training, and inference. Sampling is crucial for large graphs to create manageable subsets for efficient processing. During training, models learn from the graph's topology and node features, preparing for tasks like node classification or link prediction. Inference then applies these models to new or evolving data, adaptable to both small-scale and large-scale graph scenarios. Graph learning is vital in fields where relational data patterns play a key role, such as in social network analysis~\cite{guo2020deep, liu2022federated, yang2021consisrec} and bioinformatics~\cite{zhang2021graph,yi2022graph}.

\subsection{Programming Interfaces}
\label{sec:background:interfaces}

The diversity in graph computing also extends to programming interfaces, each tailored to specific domains within graph operations.

For graph querying, Gremlin~\cite{rodriguez2015gremlin} and Cypher~\cite{francis2018cypher} are widely used. Gremlin, part of Apache TinkerPop~\cite{tinkerpop}, offers an extensive set of operators, providing rich expressiveness for graph traversal. However, its robustness comes with complexity, as it includes over 200 steps, many with overlapping functionalities. For instance, steps like \kw{valueMap} and \kw{elementMap} both return vertex/edge properties, but with nuanced differences. This complexity poses challenges in ensuring comprehensive support within interactive graph engines.

Cypher, initiated by Neo4j~\cite{cypherneo4j}, has gained wide adoption and significantly contributed to the development of GQL~\cite{gqlstandard}, the emerging standard for querying graph databases. The increasing demand for Cypher integration into various systems, including \gs, alongside the standardization of ISO/GQL~\cite{gql}, highlights the evolving nature of graph querying interfaces. Additionally, many graph databases offer the capability to register custom stored procedures for enhanced querying functionality.

In graph analytics, following the fixed-point computation, the Pregel API~\cite{malewicz2010pregel} represents a ``think-like-a-vertex” interface~\cite{mccune2015thinking}, focusing on vertex-centric computations. PIE~\cite{fan2018parallelizing} allows for handling a partition of a \revise{graph}\eat{subgraph} as a primary element, offering an alternative methodology. On the other hand, FLASH~\cite{li2023flash} supports a flexible control flow beyond fixed-point for a wider range of algorithms.
In addition, the block-centric model~\cite{tian2013think}, edge-centric model~\cite{roy2013x}, GAS~\cite{gonzalez2012powergraph} and GraphBLAS~\cite{graphblas} are also tailored models for representing graph analytics computation.

Graph learning, especially in training and inference phases, typically employs Python-based interfaces, due to Python's prevalence in the machine learning community. This choice facilitates the integration of graph learning tasks with existing Python-based data science and machine learning ecosystems.

\subsection{Performance Requirements}
Performance requirements in real-world graph applications exhibit significant diversity, often reflecting the varied nature of the tasks.

Even within a single domain like graph querying, performance expectations can range from high query throughput (QPS) for handling multiple concurrent requests to data parallelism for complex queries typical in BI scenarios.
In online services, the emphasis often lies on high availability. Conversely, some applications might prioritize rapid processing of individual tasks, focusing on low latency and efficient single-task execution.

Graph analytics, often aligned with batch processing, also present a spectrum of performance needs. While some can be efficiently conducted in memory, others, especially those involving larger graphs with fewer machines, necessitate out-of-core processing.

Graph learning, particularly in the context of training processes, is predominantly batch-oriented. However, when dealing with large graphs, this often involves a combination of sampling and training to manage the computational demands. This process requires not just efficient data processing, but also a careful balancing of resource utilization to ensure optimal learning outcomes.

These varied performance requirements often coexist and intersect in practical scenarios, underscoring the need for diverse runtime engines and system architectures. Such diversity echoes the fundamental principle that a ``one-size-fits-all” approach is inadequate in the realm of graph computing.

\subsection{Existing Graph Computing Systems}
The complex landscape of graph computing, characterized by its diversity in storage abstractions, domain-specific applications, interfaces, and performance requirements, challenges the practicality of a ``one-size-fits-all” solution like the original design of \gs.

In this varied context, questions arise: Can we, for instance, have a system that enables high-QPS Gremlin querying on a static in-memory graph? Or set up an MPP-like Cypher process on a dynamic graph, while also accommodating GNN training on the same graph? 
Among state-of-the-art graph computing systems, specialized systems have been developed to address tailored requirements for certain types of graph workloads. For instance, various graph database systems and graph query engines~\cite{glogs, qian2021gaia, trigonakis2021adfs, jamshidi2020peregrine, su2022banyan,tigergraph,neo4j} are developed to tailor high-throughput or low-latency query execution with different underlying organization and storage of graphs. Whereas for graph analytic workloads, a different set of siloed systems~\cite{zhu2016gemini, fan2018parallelizing, chen2019powerlyra, vora2019lumos, graphx, wang2017gunrock, Ben2017Groute, gum, pandey2020c} are designed to excel in parallelizing large-scale analytical computations. To facilitate sampling-based mini-batch training, GNN systems~\cite{gandhi2021p3, yang2022gnnlab, liu2023bgl, Sun2023Legion} usually have their own graph engines and utilize in-memory storage to maximize the sampling throughputs.

\subsection{Opportunity: LEGO-like Modularity}

One possible solution to this complexity does not lie in creating siloed, specialized and fragmented systems from scratch for each unique requirement. Instead, the key is in embracing a LEGO-like modularity. This approach involves designing graph computing components -- such as various graph storages, runtime engines, workloads, and interfaces -- in a way that they can interlock or ``plug'' into each other seamlessly, much like LEGO bricks. This modularity dramatically reduces the complexity of combining different pieces, as there's no need for extensive integration or customization for each new configuration.

In essence, this modular design not only addresses the broad spectrum of graph computing requirements but also fosters innovation and adaptability in system architecture, paving the way for more efficient and effective solutions in the field of graph computing.

\section{System Overview}
\label{sec:system-overview}

\begin{figure}[t!]
    \centering
    \includegraphics[width=\linewidth]{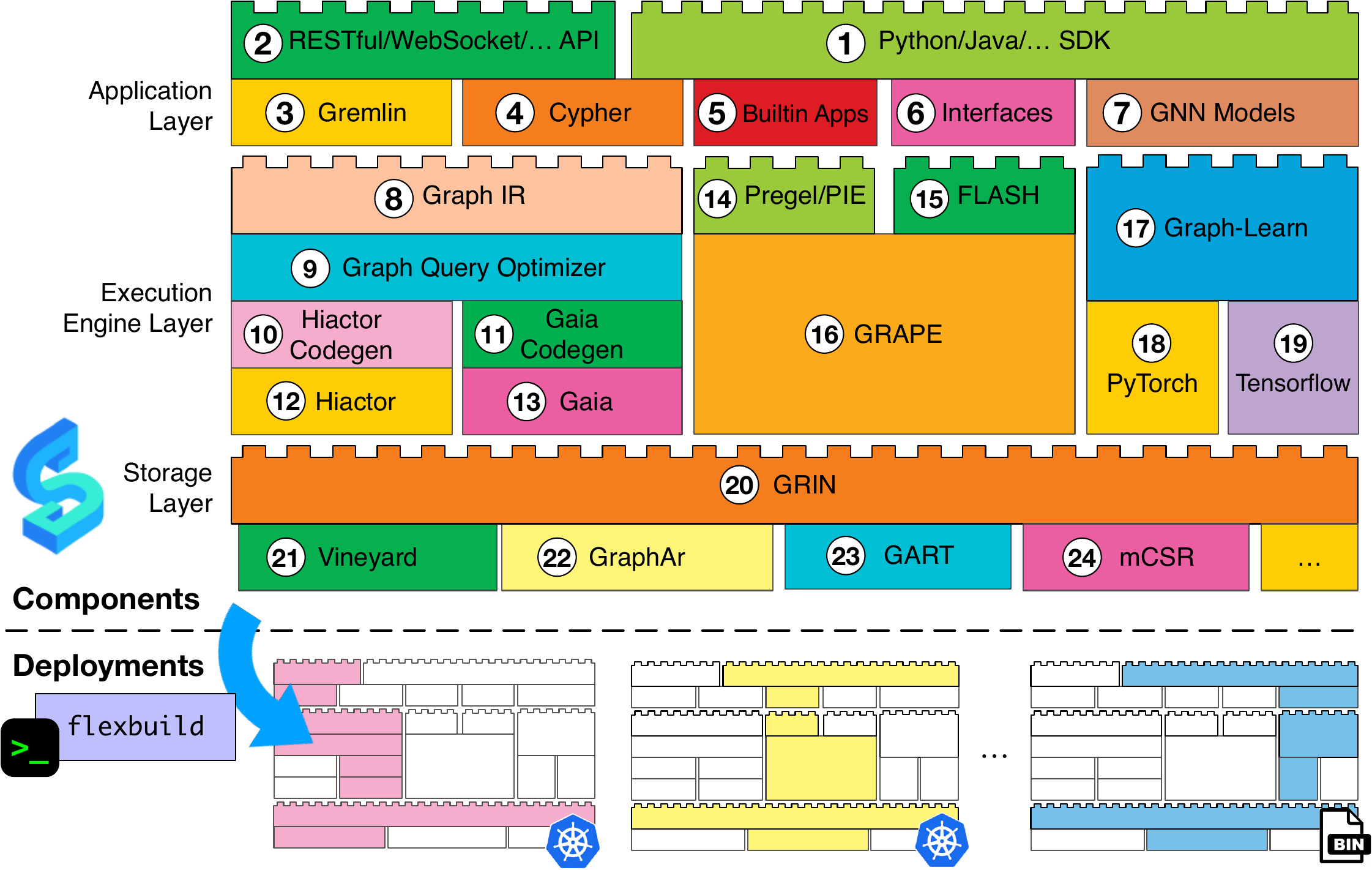}
    \vspace*{-2em}
    \caption{System architecture}
    \label{fig:architecture}
    \vspace*{-1em}
\end{figure}

% \todo{mention flexbuild}
Building on the identified opportunity, we introduce ~\gsf, 
a comprehensive system for large-scale graph processing that employs a disaggregated design. 
It features a modular architecture that reduces resource requirements 
while providing a seamless, user-friendly experience for flexible deployment. 
As depicted in Figure~\ref{fig:architecture}, 
\gsf~ comprises multiple components, 
akin to LEGO building bricks. 
Each component is designed to provide specific functionalities, 
yet some share underlying commonalities. 
This modular approach allows users to select components 
that best align with their specific requirements and build 
a tailored graph computing stack for their own scenarios.
% that users can combine like LEGO bricks to customize their graph computing stack. 

The components are classified into three layers:

% introduce 3 layers;
\stitle{Application Layer.}
Graph processing typically necessitates diverse interfaces for distinct tasks. 
~\gsf~ offers a range of user interface options, 
including SDKs (labeled by \circled{1}), 
Web Sockets, and RESTful APIs~\circled{2}. 
For graph queries, ~\gsf~ accommodates multiple query languages, 
specifically Gremlin~\circled{3} and Cypher~\circled{4}. 
For tasks involving graph analytics and learning, 
~\gsf~ features an extensive built-in library containing various 
common algorithms across different domains, 
including iterative algorithms~\circled{5}
and GNN models~\circled{7}. 
To further enhance the functionality of ~\gsf, 
it also provides interfaces~\circled{6} for the development of new algorithms.

\stitle{Execution Engine Layer.}
The components in the execution engine layer are categorized 
into three specialized groups: 
The interactive engines for graph querying and pattern matching, 
the analytical engine for graph analysis, % and iterative algorithms, 
and the learning engine dedicated to graph-based machine learning. 
A key attribute shared among these engines 
is their proficiency in efficient, 
distributed processing of queries and algorithms
on large-scale graph data.
As a query or algorithm is received, 
\gsf~ compiles it into a distributed execution plan, 
which is partitioned across multiple compute nodes for parallel processing. 
Each partition independently operates on its own compute node 
and synchronizes with other partitions via a coordinator.
This section introduces these engines, 
with more comprehensive details to be provided in Sections \ref{sec:interactive-stack} - \ref{sec:learning-stack}.

\etitle{Graph Query Engines.}
Upon receiving a query from the application layer, 
the query is parsed into a unified intermediate representation (GraphIR \circled{8}). 
This is followed by optimization through a universal Query Optimizer~\circled{9} 
and catalog module. 
The optimized logical plan employs code generation modules (~\circledd{10} and ~\circledd{11}) 
to produce the corresponding physical plan. 
Two execution engines are available, each targeting specific optimization goals. 
\hiactor~\circledd{12}, a high-concurrency engine based on the actor model, 
is optimized for high throughput. 
In contrast, \gaia~\circledd{13}, 
a \eat{distributed} dataflow-based engine, 
focuses on reducing query latency by leveraging data parallelism.

\etitle{Analytical Engine.} 
The analytical engine accommodates a variety of programming models.
This includes the widely adopted vertex-centric model Pregel~\cite{malewicz2010pregel},
the PIE model \circledd{14} based on subgraph-centric programming~\cite{fan2018parallelizing},
and the FLASH model \circledd{15} that supports non-neighbor communications~\cite{li2023flash}.
Underpinning these diverse models is \grape~\circledd{16}, 
a distributed high-performance analytical engine. 
The \grape~ engine provides a set of highly optimized core operators for 
fragment management, local evaluations on fragments, and their communication. 
Moreover, it features auto-parallelization of sequential algorithms 
for distributed environments, as well as GPU acceleration capabilities.

\etitle{Learning Engine.}
The learning engine is designated for training GNN models. 
To facilitate this, \gl~\circledd{17}~first samples 
the graph data and extracts features. 
Subsequently, these features are organized into batches and dispatched 
to a backend training engine, 
which can be either PyTorch~\circledd{18}~ or TensorFlow~\circledd{19}.

\stitle{Storage Layer.}
To address the challenges posed by diverse storage formats and data access patterns, 
\gsf~ defines a unified interface~\circledd{20}~ for graph data management and access. 
% which is the most innovative design in the storage layer. 
This interface enables seamless integration with various storage backends (~\circledd{21}-\circledd{24}) 
and makes backend complexities transparent to the execution engines. 
In Section \ref{sec:storage-layer}, we will provide further details.

\stitle{\flexbuild~ and Customized Deployments.}
To enhance user convenience, 
we introduce \flexbuild, 
a utility tool that enables users to choose specific components, 
build and generate their respective binaries or Docker images. 
These artifacts can be deployed either on a cluster or a single machine, 
allowing for customized deployments of \gsf.
For instance, 
in the real-world example described in Section \ref{sec:intro}, 
the engineers focusing on Workload 2 
might select components 
\circled{1}\circled{5}\circledd{14}\circledd{16}\circledd{20}\circledd{22}. 
Utilizing \flexbuild, 
they can build these components into a Docker image 
and deploy it on a cluster to provide a service for anti-fraud tasks. 
In contrast, 
a data scientist addressing Workload 5 
may opt for components 
\circled{2}\circled{4}\circled{8}\circled{4}\circled{9}\circledd{11}\circledd{13}\circledd{20}\circledd{23}. 
With \flexbuild, these components can be compiled 
into a binary and run on a single machine for BI analysis.
\section{Storage Layer}
\label{sec:storage-layer}

The storage layer consists of %all the heterogeneous 
a set of storage backends in \gsf,
as well as \grin, a \underline{G}raph \underline{R}etrieval \underline{IN}terface above them to provide
unified retrieval abilities for the execution layer.

\subsection{\grin, Unified Graph Retrieval Interface}
\label{sec:grin}
The \grin~ in \gsf~ is a language-agnostic interface designed to facilitate integration between diverse execution engines and storage backends. It provides comprehensive and well-defined APIs for graph retrieval, tailored to accommodate the varied graph retrieval requirements of execution engines and the distinct data models and access patterns of storage backends. This design not only simplifies the implementation of retrieval functionalities across different systems but also ensures that storage backends can clearly communicate their capabilities and limitations. The essence of \grin~lies in its ability to enable various storage backends to 'interlock' effectively with multiple engines, embodying the modular, LEGO-like approach central to the \gsf~architecture.

\begin{figure}[t!]
    \centering
    \includegraphics[width=\linewidth]{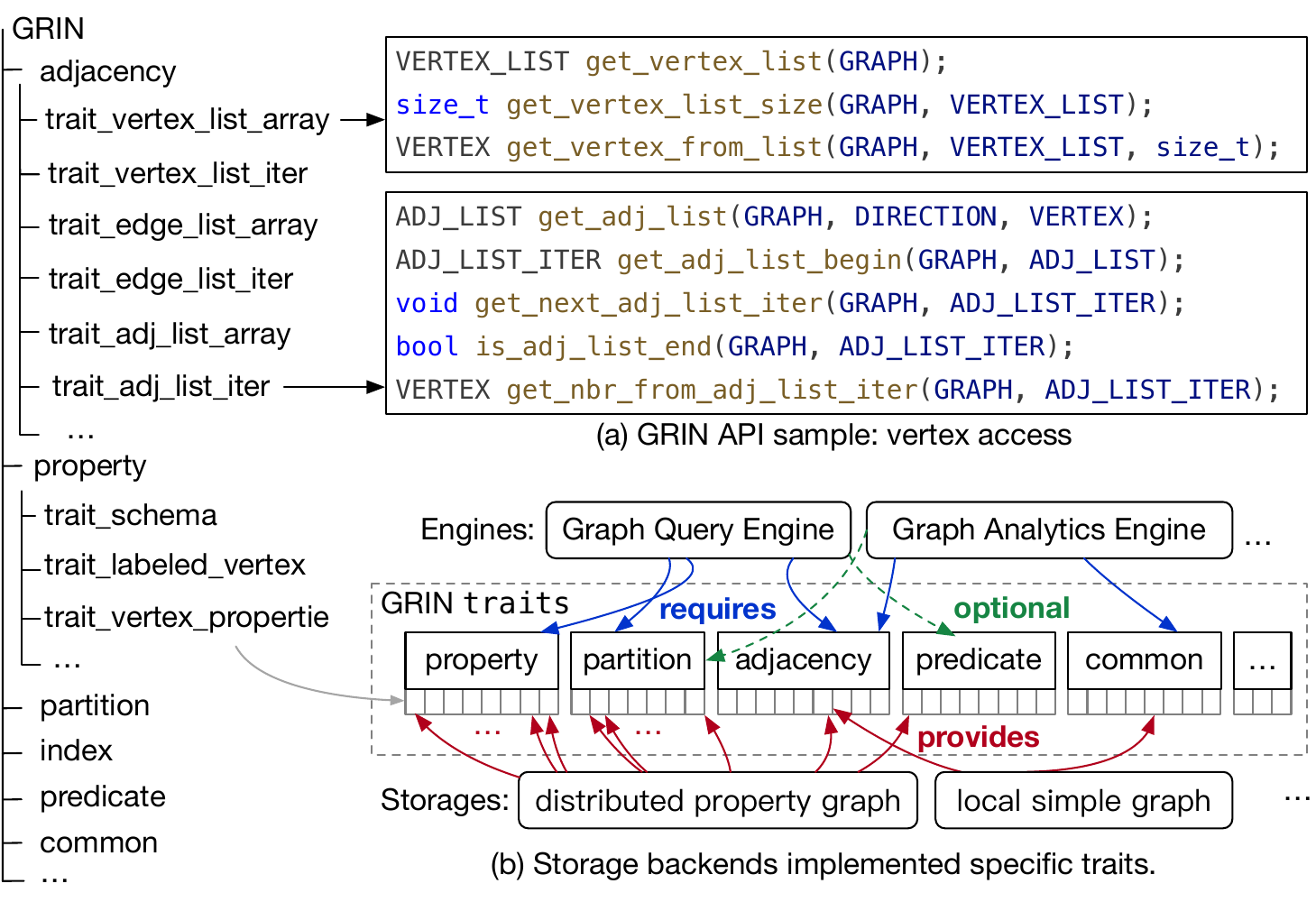}
    \vspace*{-2em}
    \caption{The design of \grin~}
    \label{fig:grin}
    % \vspace*{-2em}
\end{figure}

In particular, the \grin~ is defined in C language,
which makes it portable to systems written in other languages.
Based on a thorough analysis of the graph retrieval requirements of the execution engines in \gsf,
it abstracts the requirements into six categories as shown in the left side of Figure~\ref{fig:grin}
The \texttt{topology} category forms the core of graph abstractions, such as vertices and edges.
Based on it, the \texttt{property} and \texttt{partition} categories cover the data models of property graphs and
partitioned graphs, which are widely used in graph databases and distributed graph processing systems respectively.
The remaining three categories of \texttt{index}, \texttt{predicate} and \texttt{common} are designed 
to address the auxiliary graph operations and common system requirements,
such as indexing, predicate-pushdown and error handling.

To further characterize the graph retrieval requirements, 
% handles representing graph entities and APIs defining operations on these entities are defined 
% under several \traits~within each category.
handles (\eg \verb|ADJ_LIST|) and APIs (\eg \verb|get_adj_list(...)|) for graph entities and their operations are defined 
under several \traits~within each category.
For example, the array-like and iterator-based access are two common ways to traverse a list structure,
and the \texttt{topology} category contains corresponding \traits~for list handles, such as vertex \eat{, edge }and adjacent list.
Figure~\ref{fig:grin}(a) illustrates the APIs defined under the array-like access \trait~for
vertex list and the iterator-based access \trait~ for adjacent list.

For storage backends, they can only provide the \traits~ 
that are feasible for their own system capabilities and limitations. Similarly, 
an engine will may only require or optionally support some \traits.
In this way, a graph analytical algorithm such as PageRank will work
on a property graph, as property graph storages
will provide the \traits~for \verb|adjacency|
and \verb|partition| as shown in Figure~\ref{fig:grin}(b).

\subsection{Storage Backends}
As shown in Figure~\ref{fig:architecture}, 
\gsf~ offers various storage backends implemented \grin~APIs. 
In this subsection, 
we will introduce some of them, each tailored to specific use cases.

\stitle{\vyd.}
\label{sec:vineyard}
\vyd~\cite{vineyard} is an in-memory immutable data manager that offers out-of-the-box high-level abstraction and 
zero-copy sharing for distributed data in big data tasks, 
such as graph analytics, numerical computing and machine learning.

In \gsf, \vyd~ serves as the backend storage for in-memory graphs. 
It adopts the property graph data model and handles graph partitioning 
using edge-cut partitioning. 
In order to optimize graph retrieval, \vyd~ provides various built-in indices 
such as CSR and CSC representations for graph structures, and internal ID assignment to vertices. 
These features allow \vyd~ to effectively implement most of the \grin~\traits.

\stitle{\gart.}
\label{sec:gart} 
Graph data is not always static. 
% In many instances,
Sometimes both the graph topology and the properties of vertices/edges may be updated.
To accommodate such scenarios, \gsf~has incorporated a mutable in-memory graph storage, ~\gart~\cite{gart}, which supports multi-version concurrency control (MVCC) for dynamic graph data.

Specifically, \gart~always provides consistent snapshots of graph data (identified by a \emph{version}), 
and it updates the graph with the version number $write\_version$. 
%The computation engines can read a fresh snapshot or earlier using a version number that does not exceed $write\_version - 1$.
%After a fixed time interval, the updates in version $write\_version$ become a new consistent snapshot, and new updates are applied with the next version number $write\_version + 1$.
For read operations, a compact graph representation like CSR is the optimal choice, though it suffers from the costly overheads of write operations.
 Conversely, adjacency lists based on linked lists are efficient for write operations but perform poorly in read operations due to inadequate data locality. 
 To ensure high performance for both read and write operations, \gart~employs an efficient and mutable CSR-like data structure.
%Experimental results show that GART can provide 

\stitle{\graphar.}
\label{sec:graphar}
\graphar~\cite{GraphAr} (short for ``Graph Archive'') is a standardized file format designed for efficient storage of graph data on both local and cloud file systems.
It is developed on top of Apache ORC~\cite{ORC} and Parquet~\cite{Parquet}, two widely used columnar storage formats in the big data ecosystem.
\graphar~ serves as the default persistent format for \gsf, improving the performance of data loading and graph construction.
Additionally, it can be directly used as a data source for applications by integrating \grin.

One of the key features of \graphar~ is its ability to efficiently partition graph data into multiple data chunks, using the columnar storage feature and chunking mechanisms of ORC and Parquet.
This unique design enables it to retrieve only the relevant data chunks, potentially in parallel, eliminating the need to load the entire graph into memory before processing.
To further reduce loading overhead and improve performance, \graphar~ employs efficient encoding and compression techniques.
Furthermore, \graphar~ empowers certain graph-related operations to be executed directly at the storage layer, such as retrieving vertices with a specific label or fetching the neighbors of a given vertex, using the built-in indexes of \graphar. 
These capabilities significantly improve data management and access performance, make \graphar~ a reliable and optimized archive format for interoperability in graph computing stack.
\section{For Graph Querying}
\label{sec:interactive-stack}

Our work in \gs\cite{fan2021graphscope} introduced an interactive engine employing the Gremlin\cite{rodriguez2015gremlin} language for queries, complemented by the \gaia~\cite{qian2021gaia}
engine for their distributed execution. However, as applications burgeoned on \gs, we confronted multifaceted
challenges pertaining to both the query interface and the runtime.

As mentioned in Section~\ref{sec:background:interfaces}, users may have varying preferences for query languages, such as Gremlin and Cypher.
Additionally, the upcoming standardization
 of ISO/GQL~\cite{gql} also necessitates the development of a new query interface. Given our substantial investment in the Gremlin
 stack, including query parser and optimizer, devising a completely new compiler and runtime framework for these alternative languages
 emerges as a near infeasible task. 

\eat{
The Gremlin query language, while robust, presents a convoluted landscape with its 200+ steps,
many of which are analogously functional. For example, steps like \kw{valueMap} and
\kw{elementMap} both are used to return all vertex/edge properties, but the latter additionally
obtain the global identifier and type. Such intricacies impose a repetitive task of rendering comprehensive
support within the interactive engine. Further complicating the scenario, the pervasive adoption of Neo4j's Cypher query language~\cite{cypherneo4j}
 has spurred numerous requests for its integration into \gs. Additionally, the upcoming standardization
 of ISO/GQL~\cite{gql} also necessitates the development of a new query interface. Given our substantial investment in the Gremlin
 stack, including query parser and optimizer, devising a completely new compiler and runtime framework for these alternative languages
 emerges as a near infeasible task.
 } %eat

On the runtime front, \gaia's design, being a data-parallel, batching system, is tailored for
parallel execution of fairly intricate queries on large graphs - a natural fit for the OLAP
(online analytical processing) domain. Yet, a considerable number of applications at Alibaba
align more with the OLTP (online transactional processing) paradigm. Such applications, characterized
by small, high-concurrency queries, emphasize the necessity of addressing numerous simultaneous queries.
Given its inherent design, \gaia~ can hardly cater to the distinct needs of OLTP.

To address these challenges, we have developed a novel interactive stack of \gsf~ for graph queries. This stack integrates a
graph \emph{Intermediate Representation} (IR) abstraction, designed to capture the shared functionalities
across diverse query interfaces. Accompanying this is an optimizer anchored in equivalent transformation
rules specific to IR, along with two specialized code generators for the \gaia~\cite{qian2021gaia} and
\hiactor~\cite{hiactor} engines. These engines adeptly manage OLAP and OLTP workloads, respectively.

\begin{figure}[t!]
  \centering
  \includegraphics[width=1\linewidth]{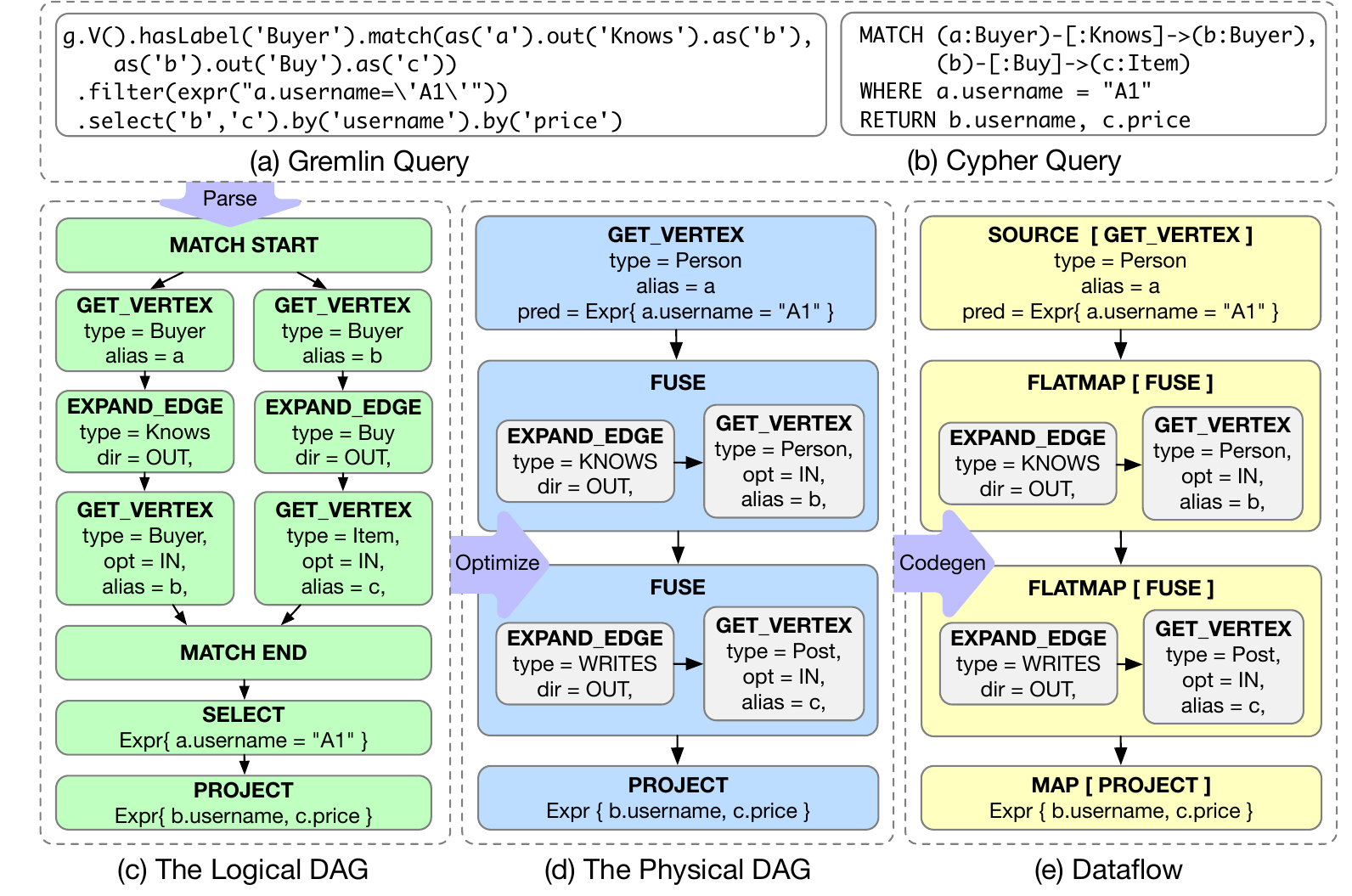}
  \vspace*{-2em}
  \caption{Compilation process of IR-based interactive stack.}
  \label{fig:gie_ir}
  % \vspace*{-2.5em}
\end{figure}

\subsection{The IR Abstraction}
\label{sec:ir}

The IR abstraction aims to encapsulate common functionalities present across various graph query interfaces.
This approach ensures that the parser, optimizer, and code generators are developed in an unified manner,
circumventing duplicate efforts. As an illustration, consider Gremlin and Cypher shown in Figure~\ref{fig:gie_ir},
both of which express the query: ``finding the purchased items' prices of friends.'' While the two queries
evidently differ in syntax (they might also diverge in query semantics~\cite{graphquerysurvey},
though exploring this further lies outside the scope of this paper), they exhibit \emph{common} functionalities
that blend graph operations with relational operations. A case in point is their shared execution of
graph pattern matching to identify items purchased by friends, followed by relational projection to
extract specific properties of interest during the search process. The evolving ISO/GQL
query language~\cite{gql} also demonstrates similar characteristics.

To address this, we devised the IR abstraction, which is tailored to semantically
encapsulate the fundamental commonalities inherent in various graph query languages. At its core,
the IR abstraction defines a data model $D$ and a set of operators $\Omega$. The model
$D$ manifests as a schema-like structure, where each data field has a String-typed name complemented
by a specific data type. Such data types can range from primitive (like Integer, Float, String) to
composite (such as List, Map), or most importantly graph-associated types. The graph-associated types
encompass Vertex, Edge, and Path, each carrying supplementary properties. For example, a \emph{Buyer}
vertex $v$, as depicted in Figure~\ref{fig:graphs}, has \emph{id} that is a unique
identifier, \emph{type} that is \emph{Buyer}, and properties of
\emph{username} and \emph{credits}.

\eat{
However, certain properties
are designated as exclusive:

\begin{itemize}[noitemsep,topsep=0pt,leftmargin=*]
\itemsep0em
\item Vertex: The `id` serves as a unique identifier, while `type` characterizes the vertex category,
evident in "Buyer" and "Item" as depicted in Figure~\ref{fig:gie_ir}.
\item Edge: `src\_vertex\_id` and `dst\_vertex\_id` pinpoint the source and destination vertices
respectively, with type specifying the edge's feature -- illustrated by "Knows" and "Purchased"
in Figure~\ref{fig:gie_ir}.
\item Path: Here, `data` embodies a collection of vertices and/or edges, whereas
`len` indicates the path's length in terms of edge count.
\end{itemize}
}

Each operator in $\Omega$ processes a set of data tuples from $D$ and yields another set of data tuples.
\eat{from $D$.} These operators can be \eat{primarily} divided into two categories: graph operators and relational operators.

\begin{itemize}[noitemsep,topsep=0pt,leftmargin=*]
\itemsep0em
\item \textbf{Graph Operators:} Central to $\Omega$ are operators specific to graph operations.
They include actions such as expanding adjacent edges from vertices (\expandedge), retrieving end vertices
from edges (\getvertex), and executing pattern matching (\matchstart, \matchend ).
\eat{References to these operators are illustrated in Figure~\ref{fig:gie_ir}(c).}
These operators are illustrated in Figure~\ref{fig:gie_ir}(c).

\item \textbf{Relational Operators:} This category emphasizes operations that project property values from vertices and edges (\project), filter vertices and edges based on specific criteria (\select),
and structure results by sequence (\order) or by categorization (\group). Such operators are
commonly used with relational databases.
\end{itemize}

For any pair of operators $o_1$ and $o_2$ within $\Omega$, a data connection can be established between
them if the output data tuples from $o_1$ align with the input requirements of $o_2$. This capability
is crucial for constructing a directed acyclic graph (DAG) that encapsulates the computational logic
of a query, often referred to as the computational DAG. This DAG will be used in two distinct stages:

\begin{itemize}[noitemsep,topsep=0pt,leftmargin=*]
\itemsep0em
\item \textbf{Logical Stage:} This is the initial stage where the query is parsed and transformed into
a semantic representation. It primarily focuses on capturing the semantic of the query,
independent of how the query will be executed.
\item \textbf{Physical Stage:} Derived from the logical stage through optimization processes, the
physical stage concretizes the execution plan for the query. It delineates the specific operations
and execution order required to fulfill the query.
\end{itemize}

Figure~\ref{fig:gie_ir}(c) illustrates the logical DAG, the semantic interpretation
of both the Gremlin and Cypher queries.
\eat{The subsequent subsection will delve into the physical
DAGs and their role in query execution.}
Next, we will delve into the physical
DAGs and their role in query execution.

\subsection{IR-based Optimizer}
\label{sec:ir_opt}

Given the logical DAG, the fundamental goal of the IR-based optimizer is to convert it into an optimized
physical DAG for efficient execution. This transformation is achieved through a combination of Rule-Based
Optimization (RBO) and Cost-Based Optimization (CBO), taking into account both semantic equivalence
and the physical context, such as the capabilities of the underlying graph store, which can be deduced
from the \grin~ interface (refer to Section\ref{sec:grin}).

Before diving into the details, we define a pattern graph as a concise graph $p$,
and the process of matching this pattern involves identifying all subgraphs within the data graph $G$
that are isomorphic to $p$.
For instance, both the Gremlin and Cypher queries in Figure~\ref{fig:gie_ir} involve matching
a pattern graph that navigates from a \emph{Buyer} vertex "a" to an \emph{Item} vertex "c" via another \emph{Buyer} vertex "b".
One of the matched instance in the data graph of Figure~\ref{fig:graphs} can be written as \emph{\{a:1, b:2, c:3\}},
where each key-value pair, such as \emph{a:1}, indicates mapping from a pattern vertex "a" to a data vertex "1".

\stitle{Rule-based Optimization.} The RBO process involves the application of a set of predefined heuristic rules to the logical DAG,
ensuring that the query's semantics remain unaltered. We highlight two rules that are most commonly applied:

\etitle{EdgeVertexFusion.} In numerous graph queries, the \expandedge~ operator is frequently
  succeeded by a \getvertex~ operator, indicating the retrieval of neighboring vertices rather than adjacent edges.
  To streamline this process, we apply the \textbf{EdgeVertexFusion} rule, which merges the \expandedge~
  and \getvertex~ operators into a unified, fused operator whenever it is pragmatically possible. The
  conditions under which these two operators can be combined vary. For example, fusion may not be feasible
  in distributed scenarios where property retrieval is required in \getvertex.
  An illustration of this
  fused operator after applying the rule is shown in Figure~\ref{fig:gie_ir}(d).
\eat{
  \item \textbf{DegreeFusion}: To obtain the degree (i.e., the number of neighbors) of a set of
  vertices $V$, the logical DAG typically involves applying the \expandedge~ operator to $V$, followed by
  a \group~ operator regarding $V$. The \textbf{DegreeFusion} rule integrates these operations when
  the degree retrieval is a native function of the underlying graph store, which is implied by the
  implementation of the \texttt{grin\_get\_adjacent\_list\_size()} interface from \grin.
  }

 \etitle{FilterPushIntoMatch.} Both the \getvertex~ and \expandedge~ operators in our design are capable
 of accepting a predicate as a parameter, enabling immediate filtering of vertices and edges upon
 retrieval from the graph store. While it is a common practice for users to apply the \select~ operation
after pattern matching, as exemplified by the queries in Figure~\ref{fig:gie_ir}, our system implements
the \textbf{FilterPushIntoMatch} rule to optimize this process. This rule actively pushes the predicate
from the \select~ operator into the pertinent graph operators.
The application of the \textbf{FilterPushIntoMatch} rule serves a dual purpose: it not only diminishes
the volume of intermediate data, enhancing performance, but also facilitates the possible downward
propagation of predicates, optimizing data retrieval at the store level. As illustrated in Figure~\ref{fig:gie_ir}(d),
the predicate \texttt{a.username = ``A1''} is effectively pushed into the \getvertex~ operator associated
with vertex "a", exemplifying this optimization in action.

\stitle{Cost-based Optimization.} We have incorporated insights from our previous research, \glogue~\cite{glogs},
to process CBO for the Interactive stack. Given that graph pattern matching is a crucial and
computationally intensive component~\cite{graphquerysurvey} of graph queries,
optimizing its execution is the main focus of \glogue. In \glogue, our approach entails tracking patterns ranging
from the smallest, single-vertex patterns to the largest, encompassing patterns with up to $k$ vertices,
along with an estimation of each pattern’s frequency. Here, the term ``frequency'' of $p$ refers to the count
of matched instances in $G$. As the execution plan for matching a pattern graph
$p$ inevitably requires the computation of various subgraphs (a subgraph is defined as a graph
comprising a subset of the original graph's vertices or edges), the cost of an execution plan can
be determined by summing the estimated frequencies of all relevant subgraphs, retrievable from \glogue.
Consequently, the execution plan with the lowest associated cost is considered optimal.

In Figure~\ref{fig:gie_ir}, the transition from the logical DAG to the physical DAG, orchestrated
by the cost-based optimization (CBO) process, results in an obvious structural transformation.
Initially, the logical DAG exhibits a bifurcated structure, which is then altered into a linear chain
in the physical DAG. This restructuring is particularly evident in the treatment of the vertices aliased
as "b". By merging these vertices in the physical DAG, we effectively eliminate the need for a separate
scanning operation for the "b"-aliased vertex, thereby reducing the associated cost.

\subsection{Code Generation}
\label{sec:codegen}
\eat{
Certainly, the physical DAG resulting from the optimization process is not directly executable.
The final stage of compilation is responsible for translating the physical DAG into executable code.
The code generation varies based on the application context, yielding code for either the \gaia~\cite{qian2021gaia}
engine in OLAP scenarios or the \hiactor~\cite{hiactor} engine in OLTP scenarios. Given their similarities,
we will concentrate our discussion on code generation for the \gaia~ engine.

The \gaia~ engine adopts a dataflow model for computation, representing it as a DAG
 analogous to our IR representation. In this model, the \gaia~ operators are general-purpose operators,
 such as \map, which processes a single data item at a time, yielding exactly one output, and \flatmap,
 which also processes single data items but may produce zero, one, or multiple outputs. Leveraging
 this semantic alignment, we can systematically map each IR operator in the physical DAG to its corresponding
 \gaia~ operator, maintaining the integral connections between operators. 
 
 This process is visually
 represented in Figure~\ref{fig:gie_ir}(e). In this process, the initial \getvertex~ operator is
 transformed into a \source~ operator, establishing the entry point for the entire computation.
 The two fused operators, designed to retrieve neighbors, naturally map to \flatmap~ operators,
 reflecting the one-to-many relationship between a vertex and its neighbors. Finally, the \project~ operator is turned into a \map~ operator, converting each vertex into the requisite property value.
}

The final step of compilation transforms the non-executable physical Directed Acyclic Graph (DAG) into executable code, varying for the \gaia~ engine in Online Analytical Processing (OLAP) or the \hiactor~ engine in Online Transaction Processing (OLTP). The \gaia~ engine, using a dataflow model similar to our Intermediate Representation (IR), processes data with operators like \map~ and \flatmap. We map each IR operator in the DAG to a corresponding \gaia~ operator, maintaining connections between operators.
This transformation is depicted in Figure~\ref{fig:gie_ir}(e). Initially, the \getvertex~ operator is converted into a \source~ operator, initiating the computation. Then, the operators intended for neighbor retrieval are mapped to \flatmap~ operators, representing the one-to-many vertex-to-neighbor relationship. Lastly, the \project~ operator is adapted into a \map~ operator, transforming each vertex into its property value.

\section{For Graph Analytics}
\label{sec:analytics-stack}

%[graph analytics workloads challenges]
% maybe move to sec-2.
\eat{
Graph analytical applications, such as community discovery and bid ranking, often present two significant challenges.
Firstly, these tasks typically require a traversal of the whole graph, dealing with large data volumes, potentially including millions or billions of vertices and edges. 
As a result, it requires the partitioning of the graph into multiple processing units for parallel processing. 
This process involves managing numerous concurrent computation and updates, as well as mitigating irregular memory access issues caused by the inherent high skewness of graph data. 
Achieving high-performance is a challenge for users.
Secondly, compared to interactive queries, analytic tasks involve more complex logic.
Currently, there is no declarative language, like Cypher, to articulate an analytic task, which demands users to manage the computational logic explicitly.
Various programming models like the Pregel~\cite{malewicz2010pregel} model and the GAS~\cite{gonzalez2012powergraph} model can be employed to describe graph analytic tasks. 
However, expressing many real-world analytic tasks, such as the Louvain algorithm and the Steiner tree algorithm, can be challenging with simple programming models. Hence, articulating a complex graph analytic task is a considerable obstacle.
}
\eat{To address these challenges, we have developed an analytic stack within \gsf.}
For graph analytics, \gsf~provides 
user-friendly interfaces 
%  by integrating various programming models for graph analytic tasks 
and a high-performance graph analytical engine \grape~\cite{fan2018parallelizing}.
% \grape~ provides highly optimized core operators. 
% By encapsulating them into more advanced programming paradigms, users can create high-performance distributed graph analysis tasks using only a few lines of code.

\stitle{User-Friendly Interfaces.}
%[built-in apps]
For the convenience of users, \gsf offers built-in algorithm packages. 
These packages feature APIs that are compatible with NetworkX~\cite{networkx}, GraphX~\cite{graphx}, and Giraph~\cite{giraph} interfaces, 
enabling users to enjoy the performance improvements offered by the software without having to modify the original code implemented in other systems.

%[interface]
If the built-in packages cannot meet user needs, 
the software provides Python/C++/Java SDKs with various programming paradigms, allowing users to employ the programming paradigm with which they are most familiar to implement complex algorithmic logic. 
For example, users can use the Pregel API to implement vertex-centric algorithms, or they can write sequential algorithms using the subgraph-centric PIE model~\cite{fan2018parallelizing}. Alternatively, users can utilize the ~\flash~ model~\cite{li2023flash}, which supports algorithms that utilize non-neighbor communication and offers great expressive capability.

\stitle{High Performance Analytical Engine.}
At the core of the analytics stack in~\gsf~ is \grape~\cite{fan2018parallelizing},
a distributed graph computing engine with the capability for auto-parallelization of sequential algorithms.
To enhance its capabilities,
we have incorporated \ingress~\cite{ingress} to facilitate algorithm auto-incrementalization,
supplementing the generality of \grape's PIE model.
\grape~ supports two types of backend to execute the graph analytic algorithms:

\etitle{The CPU backend.}% The \grape~ engine now 
~\grape~supports acceleration of graph operators with SIMD components such as AVX2 and AVX512 on the CPU backend, and emphasizes on optimizing communication and memory overhead across multiple nodes.
In terms of communication, \grape~ trades latency for throughput. It aggregates fragmented, randomly distributed small messages in memory into a continuous compact buffer before dispatching them all at once, thus enhancing bandwidth utilization. 
Furthermore, it employs \emph{varint} encoding and perfect hash to reduce peak memory usage.

\etitle{The GPU backend.}
Since many graph algorithms can significantly benefit from GPU acceleration, 
we have integrated GPU support into the \grape~ engine.
While GPUs are often less powerful than CPUs, they possess more cores and higher memory bandwidth.
For intra-GPU, \grape~integrated multiple load balance thread mapping and GPU-friendly data structure~\cite{Meng2019Gswitch} to improve the GPU utilization.
For inter-GPU, \grape~employs a work-stealing strategy 
to dynamically balance GPU workloads ~\cite{gum}. 
Idle GPU cores will steal work from busy ones to maximize GPU utilization on the fly.

\section{For Graph Learning}
\label{sec:learning-stack}

% Graph neural networks (GNNs) are widely used in real-world applications, 
% including product recommendations on e-commerce platforms and risk control in 
% financial management systems. 

% Mini-batch training (e.g., GraphSAGE~\cite{hamilton2017inductive}) is widely adopted to scale
% GNNs on large-scale graphs, which performs subgraph sampling and feature 
% collection to construct the training input for a batch.
% (1) Sampling the multi-hop subgraph for each vertex in the batch; 
% (2) Collecting features for all vertices within the sampled subgraphs. 
\eat{In industrial GNN applications, the scale of the graphs 
and the computational workloads can easily exceed the capacity of a single machine.
% Large-scale graphs are often partitioned and stored across distributed clusters, 
% necessitating distributed sampling and feature collection. 
% As for model training,
% techniques such as distributed data parallel training~\cite{pytorchddp} are 
% employed to accelerate large-scale GNN training by parallelizing the training 
% workload across multiple GPUs or machines. 
Optimizing distributed GNN training performance on large-scale graphs presents challenges in industrial scenarios. Firstly, the computational 
demands of sampling and training are asymmetrical, e.g., GPU training typically 
achieves significantly higher throughput compared to CPU-based sampling, 
resulting in a poor resource utilization and suboptimal end-to-end throughput. 
Secondly, the network and I/O overheads incurred by distributed graph
sampling, feature extraction, and the data transfer from host to device 
can substantially prolong the time required to prepare training samples.
The learning stack in ~\gsf~ is proposed to solve these challenges
by carefully parallelizing and pipelining the sampling and training processes.
}%eat
Optimizing distributed GNN training on large-scale graphs in industrial settings is challenging. First, there's an imbalance in computational demands between sampling and training. GPU training often has much higher throughput than CPU-based sampling, leading to poor resource utilization and suboptimal overall throughput. Second, the network and I/O overheads from distributed graph sampling, feature extraction, and data transfer between host and device significantly extend training sample preparation time. The learning stack in \gsf~ addresses these issues by effectively parallelizing and pipelining the sampling and training processes.

% To solve the above challenges, the learning stack in ~\gsf~
% adopts a decoupled design that physically isolates the sampling 
% and training processes, allowing them to be scaled up or out independently.
% The mechanism of asynchronous sampling and feature collection in learning stack overlaps
% the processing of I/O operations with the sampling and model training computations.
\stitle{Decoupled Sampling \& Training.}
Given the observation that the computational demands of sampling and training 
are asymmetrical, the learning stack adopts a decoupled design that physically isolates the 
sampling and training processes. This design supports independent scaling of sampling and 
training to accommodate optimal resource utilization and enhance the training throughput.
For instance, due to the high costs of GPU instances, it is often more economically 
efficient to deploy a CPU cluster for graph sampling and feature collection, while 
reserving GPU instances for training. 
% The communication between the sampling and training processes 
% are abstracted via a sample channel, which can either be a shared-memory channel or a RPC channel
% (utilizing either TCP or RDMA network), depending on the deployment configurations.

% In the case that the sampling and training processes are collocated on the 
% same physical machine, a shared-memory channel is used for inter-process 
% communication. When the sampling and training processes are deployed in separate machines
% (e.g., CPU instances for sampling and GPU instances for training), the sample
% channel is implemented as a RPC-based remote channel, utilizing either TCP or 
% RDMA network.

%%%% placing here for correct pagination.
\begin{figure*}[ht!]
    \includegraphics[width=0.95\textwidth]{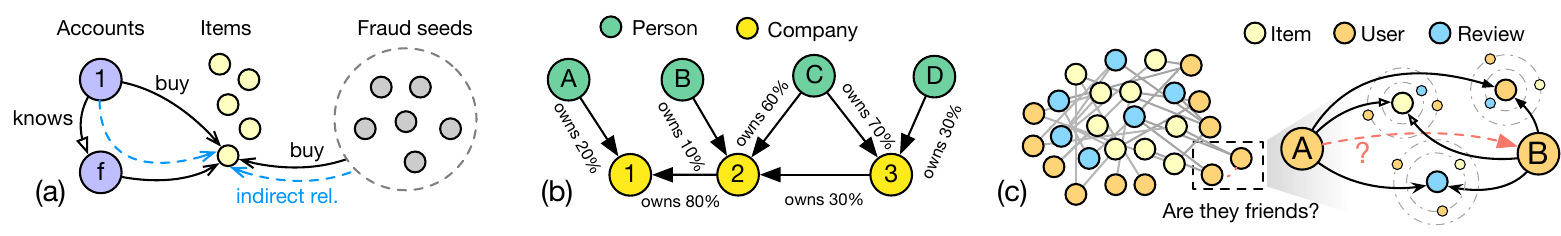}
    \vspace{-1em}
    \caption{Use cases}
    \label{fig:case-study}
    % \vspace*{-4ex}
\end{figure*}

\stitle{Asynchronous Pipelining.}
To overlap sampling computation and network communication, sampling processes 
are designed to concurrently perform subgraph sampling and feature collection 
for multiple batches. The process of multi-hop graph sampling is modeled as a dataflow, 
where each node represents a sampling hop, and the edge indicate the data dependency 
between consecutive hops. Feature collection is listed as the sink node in the dataflow. 
Since graphs are partitioned in a distributed manner, each node in the dataflow 
is parallelized across the graph partitions as distinct tasks. While awaiting the 
completion of a specific task, the sampling process schedules the execution of 
other tasks (belonging to the same or different batches) to avoid being blocked by a single task.
To prevent training processes from idling while waiting for inputs,  
a prefetch mechanism is employed to continuously retrieve data from the sample channel 
to a prefetch cache for each training process.

\stitle{Compatible with Open-source GNN Stack}
% GraphScope's current GLE engine is based on GraphLearn, with the training backend primarily using the TensorFlow 1.x series. With the rise in popularity of PyTorch, PyTorch-based GNN frameworks like DGL and PyG have accumulated a wealth of support for open-source datasets and model libraries, essentially becoming the de facto standard for GNN training frameworks. As a one-stop graph computing platform, GraphScope also needs to provide users with an experience similar to the existing open-source ecosystem, including a user-friendly interface, a rich set of models, and dataset support, while demonstrating superior performance in large-scale distributed training scenarios for graph learning tasks.
The learning stack supports both TensorFlow~\cite{tf} and PyTorch~\cite{torch} as the training 
backend. To enrich the model library that users can train with ~\gsf, 
the data-layer APIs in learning stack are designed to be compatible with Pytorch Geometric (PyG)~\cite{pyg}.
PyG models can be trained using ~\gsf~ with minimal modifications.

% The data-layer APIs in GLE are compatible with PyG~\cite{pyg}, facilitating 
% the scalable training of PyG models across distributed clusters.
% GLT can extend PyG training to distributed scenarios and leverage GPUs to accelerate graph sampling and feature extraction operations during the training process. The data-layer APIs of GLT are compatible with PyG, and only a few lines of modifications in PyG's code are needed to train PyG models using GLT.

\section{Use Cases}
\label{sec:use-cases}

To check the flexibility and effectiveness of
the modular and disaggregated design of \gsf,
we have deployed it in a variety of real-world scenarios.
Below, we delve into some of these instances to showcase the diverse deployments of \gsf.

\eat{
\stitle{~\gsf~ for BI Analysis}
In the context of Business Intelligence (BI) analysis, business analysts commonly interact with data through a Web User Interface (WebUI), conducting interactive queries and analyses. The primary concern in these scenarios is not so much concurrency, but rather achieving low latency, especially for complex queries.
Consider, for example, a BI application at Alibaba that is designed to investigate the intricate relationships between companies, their investments, and their various stakeholders. Analysts use this application to discern reputable companies and avoid less reliable ones for potential collaborations. These queries tend to be intricate, often necessitating traversals across multiple hops (typically three or more) within a vast graph data structure that contains billions of edges.
In such scenarios, achieving low latency is important, given the interactive nature of the analysts' work -- they submit queries and expect swift responses. By deploying \gs~ Insight, a BI-tailored construction of \gsf with the distributed
 \gaia engine, we have substantially enhanced performance in this use case, reducing average latency by $10\times$ compared to the preceding system. This improvement significantly enhances the user experience, fostering a more efficient and productive analytical process for the business analysts.
}

\eat{
~\gsf~ operates as an interactive query answering service over an evolving graph. In the application layer, it supports Gremlin or Cypher query languages. In the engine layer, an interactive engine (GIE) is deployed. When the engine receives a query written in Gremlin/Cypher, the compiler GAIA compiles the query into a unified interpreted representation (IR). A universal query optimizer is then applied to optimize the IR and translate it for different backends. In this BI scenario, the queries resemble OLAP style, so the IR is interpreted into a data parallel plan. Next, a code-gen module generates the physical plan and applies it to Pegasus, a distributed data-parallel compute engine based on the cyclic dataflow computation model at the core of GIE. In the storage layer, a dynamic graph store is deployed and responsible for storing the graph data. With this combination, ~\gsf~ provides a high-performance and interactive query answering service for BI analysis.
}
\eat{
\begin{figure}[ht!]
    \centering
    \includegraphics[width=0.8\linewidth]{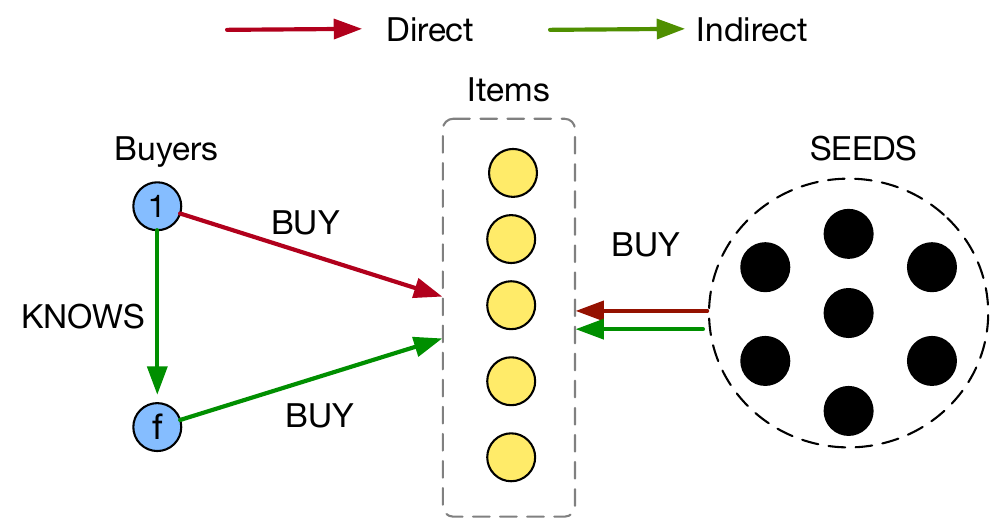}
    \caption{Real-time blacklist filter.}
    \label{fig:case1}
\end{figure}
}%eat

\stitle{Real-time Fraud Detection.}
E-commerce platforms process tens of thousands of customer orders every second. 
Among these, some may be fraudulent. For instance, certain orders might forge genuine purchasing behavior to artificially boost the popularity of specific items. As a simplified example of real-time detection,
it is to identify suspicious transactions by checking each order against \emph{fraud seeds} -- accounts previously identified with known frauds.
For an account with \texttt{id=1}, the following Cypher query checks for direct or indirect co-purchasing with \emph{fraud seeds}, as depicted in Figure~\ref{fig:case-study}(a):

\begin{lstlisting}[language=SQL, morekeywords={RETURN}, frame = none]
MATCH (v:Account{id:1})-[b1:BUY]->(:Item)<-[b2:BUY]-(s:Account)
WHERE s.id IN SEEDS AND b1.date-b2.date < 5  /*within 5 days*/
        WITH v, COUNT(s) AS cnt1
MATCH (v)-[:KNOWS]-(f:Account), (f)-[b1:BUY]->(:Item)<-[b2:BUY]-(s:Account) WHERE s.id IN SEEDS WITH v, cnt1, COUNT(s) AS cnt2
WHERE w1 * cnt1 + w2 *  cnt2 > threshold 
RETURN v
\end{lstlisting}

To facilitate such queries, we deploy \gsf~ for OLTP graph queries,
utilizing \hiactor~ as the computing engine, and
\gart~ as the storage engine. When an order is placed, %it's first recorded in a relational DBMS. Simultaneously, 
an \emph{(Account)-[Buy]-(Item)} edge is added to \gart, leveraging its dynamic graph storage capability. This query is executed in \gsf. 
If the query returns any records, indicating that the weighted average of the number of direct and indirect relationships exceeds a predefined \texttt{threshold}, an alert will be triggered. This preemptive step is vital to prevent direct lodging of potentially fraudulent orders. %Subsequent investigations can then be conducted to reduce false alarms.

%The deployment has demonstrated exceptional performance, setting new records in the recent LDBC Social Network Benchmark~\cite{ldbc}. As a result, \gs~Interactive has distinguished itself as the graph system capable of meeting Alibaba's stringent business requirements.
\eat{
\begin{figure}
    \centering
    \includegraphics[width=0.8\linewidth]{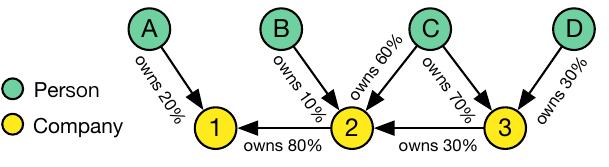}
    \caption{Equity analysis}
    \label{fig:equity-analysis}
\end{figure}
}%eat

\stitle{Equity Analysis.}
In the context of financial analysis,
analysts frequently aim to identify the dominant shareholders
responsible for steering a company,
typically those who cumulatively hold more than 51\% of the company's shares.
It is non-trivial to find the answer in real-world scenarios.
For instance, the Company 1 in Figure~\ref{fig:case-study}(b) is owned by
Person A alongside a sequence of corporate shareholders.
The objective for analysts is to ascertain the genuine controller of Company 1,
\ie Person C, who holds 51\% of the shares: $0.8\times0.6$ through Company 2
and $0.8\times0.3\times0.7$ via Company 3.

This task can be tackled using a \gsf~ deployment equipped with analytical modules.
In graph modeling,
both shareholders and companies are represented as vertices,
while investments are depicted as edges.
Each edge carries a weight, indicating the share percentage.
Within the application layer,
users opt to implement a modified label propagation algorithm with the GraphX API.
This algorithm calculates the shares for each company during each propagation iteration.
Subsequently, the algorithm is bundled into a \emph{jar} and executed on \grape,
the high-performance analytical engine in the engine layer of \gsf.
In the storage layer,
\grin~ selects the in-memory graph store \vyd~ as its preferred storage backend to managing
the rarely modified graph data.

% ~\gsf~ is both performant and easy-to-use for graph analytical jobs.
% In this scenario, applications are provided as iterative algorithms in
% the built-in library or customized by the users.
% The algorithms can be implemented using FLASH-style API,
% Pregel API, or GRAPE native API.
% All these types of algorithms are compiled in the engine layer and run on GRAPE,
% a processing system proposed in this paper published in SIGMOD 2017.
% GRAPE distinguishes itself from prior systems by its ability to parallelize sequential
% graph algorithms as a whole. In GRAPE, sequential algorithms can be easily “plugged into”
% with only minor changes, allowing them to efficiently handle large graphs.
%  To achieve high performance, deployments for this scenario typically
%  choose an in-memory graph store for the storage layer.

\stitle{Social Relation Prediction.}
\eat{Social relations of users are essential for e-commerce platforms,
as they offer insights into user behavior and preferences,
enabling targeted marketing and personalized recommendations.
However, compared to social network platforms, the user relations on
e-commerce platforms often do not include the users' complete
social network. The link prediction task in GNNs (e.g., NCN~\cite{wang2023neural})
focuses on predicting the likelihood of an unobserved relationship existing
between two vertices, and thus can be utilized to enrich the social
relations on e-commerce platforms.
As shown in Figure~\ref{fig:case-study}(c), the sampling phase of NCN
requires extracting the first-order common neighbors for both
vertices of every training edge and then conducting k-hop subgraph sampling
for each common neighbor vertex.
This process presents a significant challenge due to its computational
intensity and complexity, especially when dealing with large-scale graphs.
}
Social relations on e-commerce platforms are crucial,
as they provide insights into user behavior and preferences,
facilitating targeted marketing and personalized recommendations.
However, these platforms often lack a complete social network of users, unlike social network platforms.
Graph Neural Networks (GNNs), such as NCN~\cite{wang2023neural}, are employed to predict potential unobserved relationships between users, thereby enriching the social connections on e-commerce sites.
The NCN's sampling phase, depicted in Figure~\ref{fig:case-study}(c), involves extracting first-order common neighbors for each training edge's vertices and performing k-hop subgraph sampling around each common neighbor. This task is particularly challenging in terms of computational demand and complexity, especially for large-scale graphs.
% GNNs are extensively used in real-world applications. The link prediction
% task in GNNs involves predicting the likelihood of a relationship
% existing between two nodes in a graph. In e-commerce platforms, by modeling
% the relationships between users and items as a graph, GNNs can be used to
% make product recommendations for users. Consider an e-commerce graph that
% consists of two types of edges: ``(User)-[Buy]-(Item)''
% (denoting the purchase action of a user), and ``(Item)-[To]-(Item)''
% (indicating that two items coexist in the purchase history of a user).
% We can train a GraphSAGE model on the ``(User)-[Buy]-(Item)''  edge to
% predict the likelihood that a user will buy a specific item. Specifically,
% the link prediction task is treated as a binary classification task. For
% each trained edge, multi-hop sampling will be conducted on both the ``User''
% and ``Item'' nodes. The sampled subgraphs are then used to train and update the model.

The learning stack in ~\gsf~ can efficiently support training
NCN on billion-scale graphs. Depending on the graph scale, the sampling
servers in ~\gl~ can be flexibly scaled out to enhance the sampling throughput.
To accommodate heterogeneous hardware and improve resource utilization in
sampling servers, the learning stack also supports configuring the sampling
devices (CPU or GPU) and the concurrency of sampling processes in each
server. Training servers asynchronously pull the sampling results from sampling servers
and can be scaled to match the sampling throughput. As the original social relation
graph remains unchanged and will be frequently accessed during training, ~\vyd~ is
selected as the storage backend due to its I/O efficiency. Graph data in
~\vyd~ are accessed by sampling processes through ~\grin.

% In e-commerce platforms like Taobao, the user-item graph could consist of billions of nodes and trillions of edges. The learning stack in ~\gsf~ is designed to efficiently support the training of GNNs on such large-scale graphs. The sampling and training workers within the learning stack can be flexibly scaled out according to the size of the graph and the complexity of the model, thereby maximizing end-to-end throughput. Additionally, as the learning stack in ~\gsf~ is compatible with PyG, link prediction models in PyG can be trained in a distributed manner using ~\gsf~.

% GNNs have been effectively utilized in numerous real-world applications, such as e-commerce recommendation systems and financial risk control platforms, where billion-scale graphs are prevalent. The learning engine in ~\gsf~ (GLE) efficiently supports distributed GNN training on large-scale graphs in these industrial scenarios. GLE offers both Python and C++ interfaces for graph sampling operations and a Gremlin-like GSL (Graph Sampling Language) interface to simplify the definition of sampling queries. For GNN models, GLE provides a variety of paradigms and processes for model development, and also includes a rich set of example models. Users can flexibly choose TensorFlow or PyTorch as the training backend. To enable a flexible and cost-effective resource configuration, the processes of sampling and training in GLE are decoupled, allowing each to be independently scaled to achieve the best end-to-end throughput.

\stitle{Cybersecurity Monitoring.}
Cybersecurity represents a paramount
concern for numerous enterprises,
with a particular emphasis on
thwarting Trojan attacks.
We detailed the task and its Gremlin-based solution
in our previous work~\cite{fan2021graphscope}.
With \flexbuild,
users can effortlessly select pertinent components from \gsf~
to construct a tailored graph BI stack optimized
for this specific task.

\section{Evaluations}
\label{sec-perf}

In this section, we evaluate ~\gsf's capability of
efficient processing of large graphs for both synthetic workloads
and the real-world applications described in Section \ref{sec:use-cases}.

\subsection{Synthetic Workloads}

Using synthetic workloads,
we assessed the performance of the storage layer and various deployments on \gsf.
The datasets and their abbreviations are listed in Table~\ref{table:datasets}.
If not otherwise mentioned, all experiments were conducted on a managed K8s
cluster consisting of 8 nodes. Each node was equipped with dual 26-core
Intel(R) Xeon(R) Platinum CPUs at 2.50GHz and 768 GB of memory.
The nodes were interconnected via a 50 Gbps network.

%%% ------------- Exp-3 BEGIN ------------- %%%
\begin{table}[htbp]
  \fontsize{8}{9}\selectfont
  \centering
  \caption{Datasets used in synthetic workloads.}
  \vspace{-2.5ex}
  \begin{tabular}{c|l|c|c}
    \toprule
    \textbf{Abbr.} & \textbf{Dataset}&\textbf{|V|}&\textbf{|E|}\\
    \midrule
    \texttt{FB0}  & datagen-9\_0-fb~\cite{ldbc-analytics} & 12.8M  & 1.05B \\
    \texttt{FB1}  & datagen-9\_1-fb~\cite{ldbc-analytics} & 16.1M  & 1.34B \\
    \texttt{ZF}   & datagen-9\_2-zf~\cite{ldbc-analytics} & 434.9M & 1.04B \\
    \texttt{G500} & graph500-26~\cite{ldbc-analytics}     & 32M    & 1.05B \\
    \texttt{WB}   & webbase-2001~\cite{florida-spmat-datasets}    & 118M   & 1.71B \\
    \texttt{UK}   & uk-2005~\cite{florida-spmat-datasets}         & 39.5M  & 1.57B \\
    \texttt{CF}   & com-friendster~\cite{snapnets}   & 65.6.5M  & 1.81B \\
    \texttt{TW}   & twitter-2010~\cite{snapnets}         & 41.7M  & 1.47B \\
    \texttt{IT}   & it-2004~\cite{florida-spmat-datasets}         & 41M    & 1.15B \\
    \texttt{AR}   & arabic-2005~\cite{florida-spmat-datasets}     & 22.7M  & 1.11B \\
    \texttt{PD}   & ogbn-products~\cite{ogbn}   & 2.4M   & 62M   \\
    \texttt{PA}   & ogbn-papers100M~\cite{ogbn} & 111M   & 1.6B  \\
    \hline
  \texttt{SNB-30}  & \multirow{3}{*}{ \begin{minipage}{0.5\linewidth}Datasets (-x for scale factor x) generated for LDBC social network benchmark~\cite{ldbc-snb} \end{minipage}} & 89M &  541M \\
  \texttt{SNB-300} &   & 817M & 5.27B  \\
  \texttt{SNB-1000} &  & 2.69B  & 17.79B  \\
  \bottomrule
  \end{tabular}
  % \raggedright
  \label{table:datasets}
  \end{table}

\stitle{Exp-1. Storage Performance.}
We evaluate the effectiveness of \grin~ and test the performance of \gart~and \graphar, which are the newly introduced storage backends in \gsf.

Firstly, to showcase effectiveness of \grin~ which makes backend complexities transparent to the execution engines,
we conduct experiments on three applications with varying storage backends.

\begin{figure*}[tb!]
  \vspace{-4ex}
  \begin{center}
  \centerline{\includegraphics[scale=0.5]{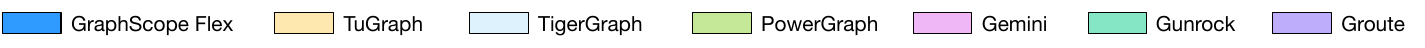}}
  \vspace{-5ex}
  \end{center}
  \begin{center}
  \setlength{\subfigcapskip}{-5pt}

  %%%%%%%%%%%%%%%%% Exp-1: GRIN and storage backends %%%%%%%%%
  %%%%%Fig 1:
  \subfigure[\small{\grin~with backends}]{\label{fig:grin-33}
  {\includegraphics[height=3cm, width=4cm]{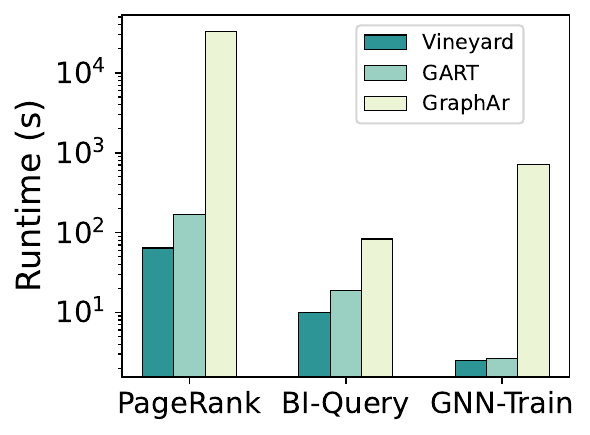}}}
  % \vspace{-1.5ex}
  %%%%%Fig 2:
  \subfigure[\small{\grin~overhead}]{\label{fig:grin-v6d-wo}
    {\includegraphics[height=3cm, width=4cm]{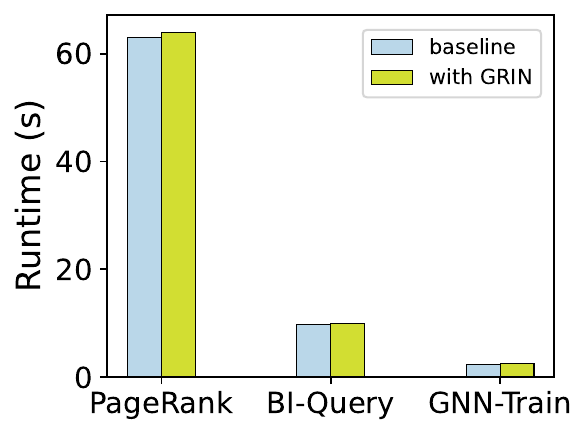}}}
  \vspace*{-1.1ex}
  %%%%%Fig 3:
  \subfigure[\small{Read performance of \gart}]{\label{fig:exp-gart}
  {\includegraphics[height=3cm, width=4cm]{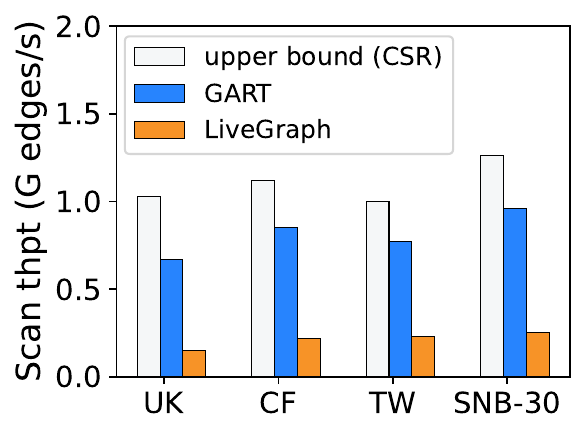}}}
  \vspace*{-1.1ex}
  %%%%%Fig 4:
  \subfigure[\small{Loading speedup of \graphar}]{\label{fig:exp-graphar}
  {\includegraphics[height=3cm, width=4cm]{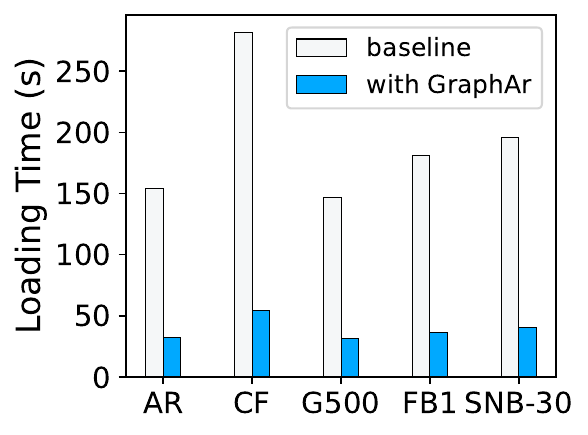}}}

  %%%%%%%%%%%%%%%%% Exp-2: interactive queries %%%%%%%%%
  \subfigure[\small{Query optimization}]{\label{fig:exp-opt}
  {\includegraphics[height=3cm, width=8cm]{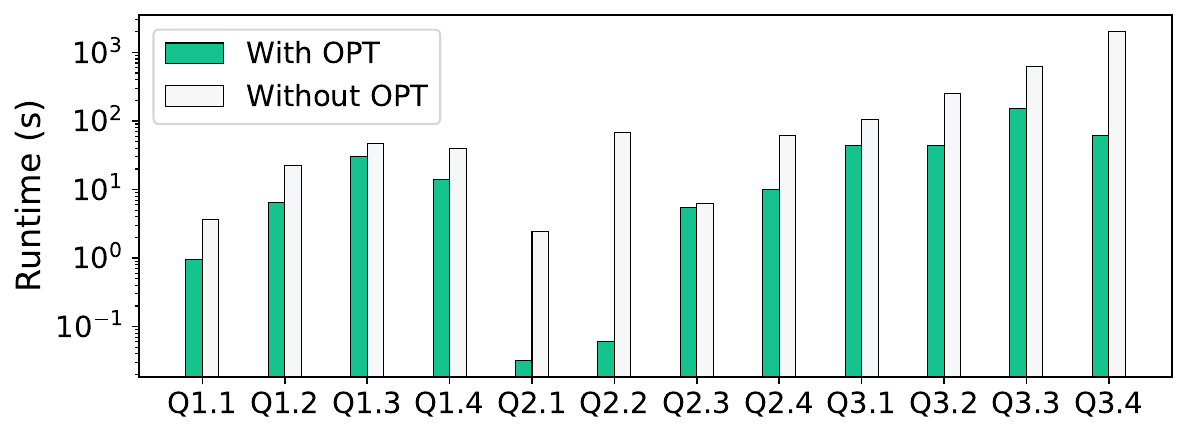}}}
  \vspace{-2.5ex}
  \subfigure[\small{OLTP-like queries}]{\label{fig:exp-ia-perf}
  {\includegraphics[height=3cm, width=8cm]{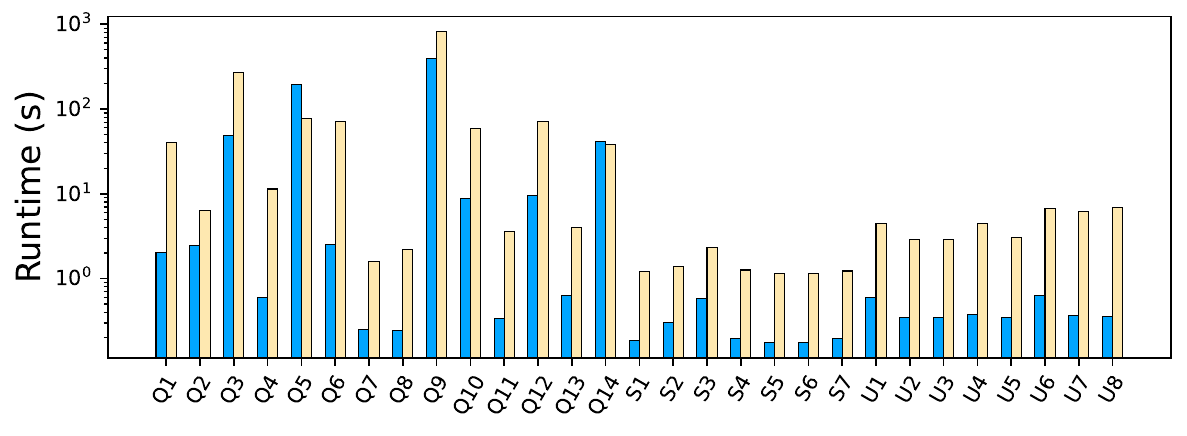}}}
  \subfigure[\small{OLAP-like queries}]{\label{fig:exp-bi-perf}
  {\includegraphics[height=3cm, width=8cm]{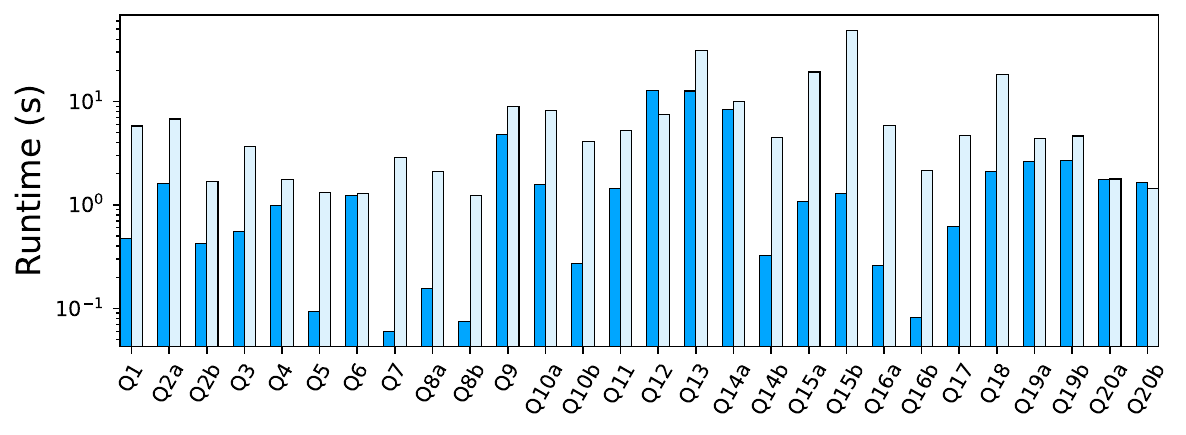}}}
  \vspace{-1.5ex}
  \subfigure[\small{PageRank on CPUs}]{\label{fig:exp-gae-cpu-pagerank}
  {\includegraphics[height=3cm, width=4cm]{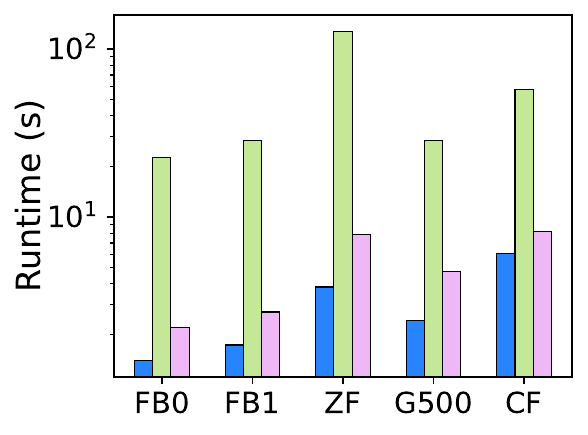}}}
  \vspace*{-1.1ex}
  \subfigure[\small{BFS on CPUs}]{\label{fig:exp-gae-cpu-bfs}
  {\includegraphics[height=3cm, width=4cm]{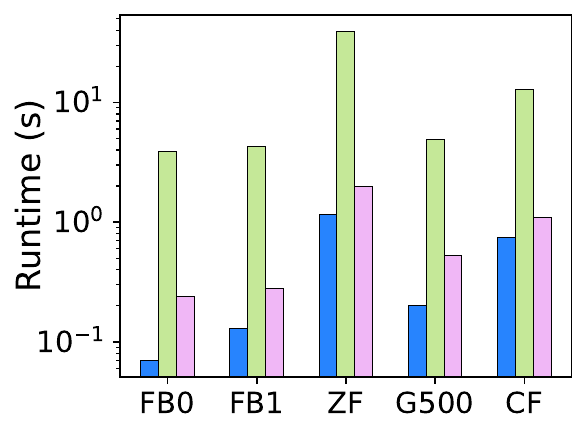}}}
  % \vspace*{-1.1ex}

  \subfigure[\small{PageRank on GPUs}]{\label{fig:exp-gae-gpu-pagerank}
  {\includegraphics[height=3cm, width=4cm]{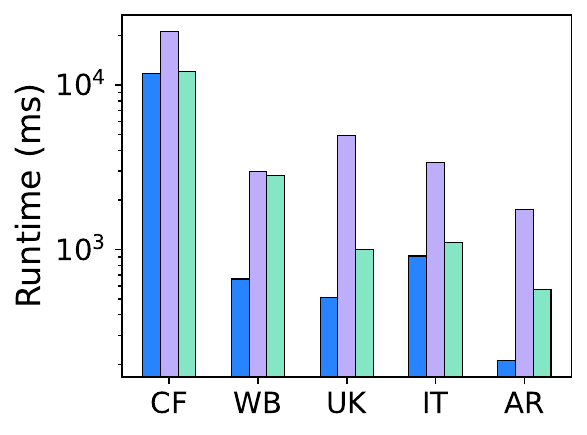}}}
  %\vspace{-1ex}
  \vspace*{-1.1ex}
  %%%%%Fig 3: BFS
  \subfigure[\small{BFS on GPUs}]{\label{fig:exp-gae-gpu-bfs}
  {\includegraphics[height=3cm, width=4cm]{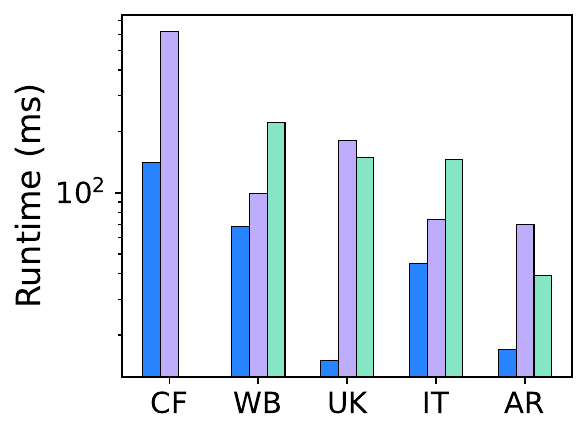}}}
  \subfigure[\small{Sampling, scale up.}]{\label{fig:gle-scaleup}
  % \caption{Scale Up}
  {\includegraphics[height=3cm, width=4cm]{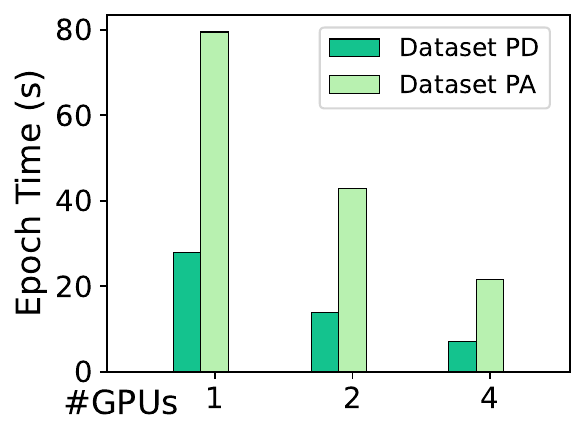}}}
  \subfigure[\small{Sampling, scale out.}]{\label{fig:gle-scaleout}
  % \caption{Scale Out}
  {\includegraphics[height=3cm, width=4cm]{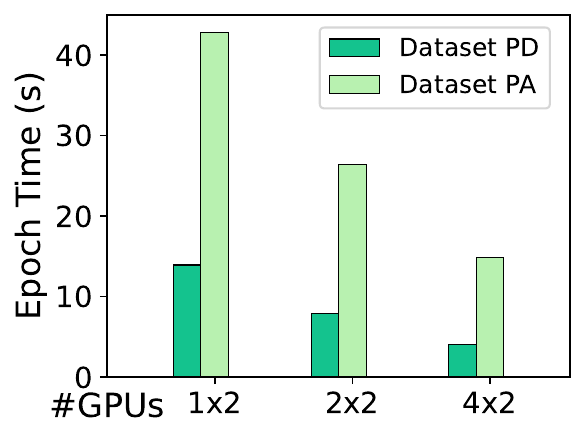}}}
  % \vspace{-1.1ex}
  \end{center}
  \vspace{-3ex}
  \caption{Performance evaluation}
  \label{fig:performance}
  \vspace{-2ex}
  \end{figure*}

The applications are PageRank(on \texttt{CF}), BI-Querying(on \texttt{SNB-30}) and GNN-Training(on \texttt{PD}) which are typical workloads
of the graph analytics, interactive query and graph learning respectively, 
while the backends are \vyd, \gart, and \graphar~which are the in-memory immutable store, in-memory dynamic store
and external storage respectively.
The execution engines use the \grin~APIs to access the graph data,
so each application is implemented only once and can be deployed on different storage backends.
We report the execution time of PageRank, average querying time of BI queries and one-batch training time of GNN in Figure~\ref{fig:grin-33}.
We can see that all the combinations can generate correct results in a reasonable time.
\eat{In general, \vyd~ is the fastest backend, because it is an in-memory immutable store.
\gart~ is slower than \vyd, 
since its architecture is more complex to handle dynamic updates.
Although \graphar~ is the slowest, its design for archiving results in additional I/O overheads when used for direct data retrieval.}
Generally, \vyd~ is the fastest backend due to its in-memory and immutable design. \gart~ is slower, as its more complex architecture accommodates dynamic updates. \graphar, being the slowest, incurs extra I/O overheads for direct data retrieval due to its archiving-focused design.

To further demonstrate the effectiveness of \grin, we compare the performance of \gsf~ with the baseline
(\ie the original \gs~ without \grin). In \gs, the execution engines are tightly coupled with the 
only default storage backend \vyd, so that the graph retrieval implementations are specifically optimized for \vyd~ and cannot be easily extended to support other backends.
As shown in Figure~\ref{fig:grin-v6d-wo},
one may find that the performance of \gsf~ with \grin~ is comparable to baseline for all three applications,
with only a slight overhead less than $8\%$.
This means that \grin~is well designed and does not introduce significant overheads to \gsf.

Next, we demonstrate the efficiency of \gart, which provides high performance read operations on graph data while allowing updates to graph data.
To see this, we compared the read performance (measured by edge scan throughput) with a state-of-the-art dynamic graph storage LiveGraph~\cite{zhu13livegraph} and a \emph{static} graph storage CSR, on four datasets (\texttt{UK},  \texttt{CF}, \texttt{TW} and \texttt{SNB-30}).
Note that the performance of CSR is the upper bound of a dynamic graph storage as it assumes the graph is immutable.
As shown in Figure~\ref{fig:exp-gart}, on average \gart~outperforms LiveGraph $3.88\times$ on read performance, and can achieve $73.5\%$ throughput compared with static graph storage.

% Lastly, we illustrate the graph construction efficiency of \graphar,
Lastly, we compare the performance of building graphs from external storages 
in the format of \graphar~with the baseline,
where the datasets are in CSV format.
The results are shown in Figure~\ref{fig:exp-graphar}.
We can see that \graphar~can significantly improve the performance of graph construction,
with a speedup of around $5\times$ for all datasets.

% \begin{itemize}
%     \item fig-1, graph retrieval performance(y), x: w/o GRIN, many backends.
%     \item fig-2, various graph computing performance(y) with/without GRIN, x: applications, each with two bars.
%     \item fig-3, application on relational data performance(y), using GART or not, x: applications, each with two bars;
%     \item fig-4, graph construction performance(y), with/without GraphAr(or adapters), x: datasets, each with two bars(or more);
% \end{itemize}

\stitle{Exp-2. Graph Query Performance.}
We evaluate the performance of \gsf~ for processing graph queries,
demonstrating the effectiveness of the query optimization including RBO and CBO (Section~\ref{sec:interactive-stack}), and
the efficiency of the deployments of \gsf~ for handing OLTP and OLAP queries, respectively.

To evaluate the effectiveness of query optimization in \gsf, we generated three distinct sets of queries,
which are referred to as Q1, Q2, and Q3 as given in \cite{opt-queries}. Each set comprises four queries designed to
specifically test different optimization strategies: the \emph{EdgeVertexFusion} and \emph{FilterPushIntoMatch} rules
in RBO, and the CBO strategy. 
\eat{These queries were intentionally designed to reflect the impacts of the respective optimization techniques.
For the tests, we deployed \gsf~ for graph queries, employing \gaia~ as the computing engine and the \vyd~ storage for microbenchmarking.}
 The experiments were conducted on a single node of the cluster, \kwlog
 \eat{executing each query with 52 threads. Each query was run 20 times to calculate the average runtime.}
 \eat{We set a timeout limit of 60 seconds; queries exceeding this limit were assigned a default runtime of 60 seconds.
 Figure~\ref{fig:performance}(e) shows the RBO optimization archives remarkable performance improvements: the \emph{EdgeVertexFusion} rule yielded an average speedup of $2.9\times$,
while the \emph{FilterPushIntoMatch} rule achieved an impressive average speedup of $279\times$.}
Figure~\ref{fig:performance}(e) shows the performance gain with RBO optimization: the \emph{EdgeVertexFusion} rule yielded an average speedup of $2.9\times$, and the \emph{FilterPushIntoMatch} rule achieved an impressive average speedup of $279\times$.
Similarly, queries optimized with CBO performed 11$\times$ better compared to those without CBO.

\eat{We proceeded to evaluate the \eat{end-to-end }performance of \gsf~ in processing OLTP-like queries.
For this purpose, we present the official audit results from the LDBC Social Network Interactive (SNB-IA) benchmark~\cite{snb-interactive}.
}
For the performance of \gsf~ in processing OLTP-like queries, we present the official audit results from the LDBC Social Network Interactive Benchmark~\cite{snb-interactive}.
We deployed \gsf~ as the OLTP graph querying stack, 
with \hiactor~ serving as the computing engine.
% a specific configuration for processing high-throughput, concurrent OLTP queries.
To align with the existing reports in \cite{ldbc-ia-audit},
the benchmark was conducted on a single
machine with a CPU of 24 cores and 384GB of memory.
Figure~\ref{fig:exp-ia-perf} reports the auditing results~\cite{ldbc-ia-audit} of
\gsf~ against TuGraph on the \texttt{SNB-300} dataset,
displaying the average latency for 14 complex queries (C1-C14), 7 short queries (S1-S7), and
8 update queries (U1-U8) in the benchmark.
% In terms of latency,
The results show \gsf~ outperforms TuGraph in all queries except C5, achieving an average speedup of 8.92$\times$.
Furthermore, the recorded throughput for \gsf~ is 33,261 ops/s, which is 2.45$\times$ higher than TuGraph's \eat{throughput of }13,532 ops/s.

To assess the performance of \gsf~ in handling OLAP-like queries, 
we utilized the LDBC Social Network Benchmark
Business Intelligence (SNB-BI) workloads~\cite{snb-bi}.
% which is well-suited for these types of workloads.
In this experiment, 
we configured \gsf~ with \gaia~ as the distributed computing engine,
as outlined in Section~\ref{sec:interactive-stack} that is tailored for OLAP queries. We conducted the SNB-BI benchmark evaluation
on the \texttt{SNB-1000} dataset of the first snapshot that encompasses over 99\% of the total data, executing a total of 20 queries (some queries having different variants) with the frequency and parameters specifying by official auditing.
As comparison, we gathered auditing results from TigerGraph~\cite{tiger-audit}, the state-of-the-art performance of the SNB-BI benchmark.
It is important to emphasize that while our tests were not subject to official auditing, 
the comprehensiveness of our evaluation lends credibility to the results.
\eat{We focused on the first snapshot of the dataset, which encompasses over 99\% of the total data.
This extensive coverage, combined with our adherence to the official auditing procedures (specifying the order,
frequency, and parameters for all queries) ensures that our results are comparable to those obtained from official audits.}
%We focused on the dataset's first snapshot, which encompasses over 99\% of the total data, and adhered to official auditing procedures by specifying the order, frequency, and parameters for all queries.
Our tests were conducted on a cluster of 4 machines that matched TigerGraph's settings, including
64 CPU cores (128 threads), and a total memory of 1.5TB.
The average latency results for all queries are displayed in Figure~\ref{fig:exp-bi-perf}. With the exception of query 12 and 20b, \gsf~ \eat{consistently}
outperforms TigerGraph by an average speedup of 10$\times$.

% \begin{figure}[htbp]
% \centering
% \includegraphics[width=3.4in]{./exp/cpu-gae-perf.pdf}
%     \caption{A performance comparison of \gsf CPU backend, Gemini~\cite{zhu2016gemini}, GraphX~\cite{graphx}, and Giraph~\cite{giraph}.}
% \label{fig:cpu-gae-perf}
% \end{figure}

% \begin{figure}[htbp]
% \centering
% \includegraphics[width=3.4in]{./exp/gpu-gae-perf.pdf}
%     \caption{A performance comparison of \gsf GPU backend, Gunrock~\cite{wang2017gunrock}, and Groute~\cite{Ben2017Groute}.}
% \label{fig:gpu-gae-perf}
% \end{figure}

\stitle{Exp-3. Graph Analytics Performance.}
% To test the performance of analytical application of \gsf, we evaluate \gsf and its competitors to two typical graph algorithms to benchmark: PageRank (PR), and single-source shortest path (SSSP) algorithms.
To test the analytics performance, we conducted evaluations using the LDBC Graphalytics Benchmarks~\cite{ldbc-analytics}, comparing ~\gsf~ with the state-of-the-art CPU-based and GPU-based graph processing systems~\cite{gs-ldbc-analysis-report}.
We report the performance of PageRank and breadth-first search (BFS) algorithms here, \kwlog
% The evaluated graphs are shown in Table~\ref{table:gae-datasets}.%\looseness=-1
\eat{
According to the latest results, Gemini~\cite{zhu2016gemini}, PowerGraph~\cite{gonzalez2012powergraph} targeted for CPU-baed graph processing and Gunrock~\cite{Wang2015Gunrock} and Groute~\cite{Ben2017Groute} targeted for GPU-baed graph processing delivered relatively better performance than the others owing to its continuous evolution. Thus we compare \gsf with these systems.
We ran CPU test on a single node with Intel(R) Xeon(R) Platinum 8269CY CPU equipped with 96 GB host memory and GPU test on a single-node with 8 NVIDIA V100 GPUs with 32 GB of global memory.
%We compiled all the GPU programs using NVIDIA's \texttt{nvcc} compiler (version 10.1.0).
The measured results in all experiments ignore the graph loading time.
}
Figures~\ref{fig:exp-gae-cpu-pagerank} to ~\ref{fig:exp-gae-gpu-bfs} display the performance of these algorithms across different systems and datasets, with \gsf~ consistently outperforming others.
% Figures~\ref{fig:exp-gae-cpu-pagerank} to ~\ref{fig:exp-gae-gpu-bfs} show the elapsed time for these algorithms in different systems and datasets.
% For both CPU and GPU backend and all datasets used,
% \gsf~ outperforms others consistently.
%  graph processing systems, and can process large graphs that other systems failed to process.
Compared with the CPU-based systems, it is on average 25.1$\times$ (resp. 2.3$\times$) faster than PowerGraph (resp. Gemini), up to 55.7$\times$ (resp. 3.4$\times$).
For GPU-based systems, it is on average 3.3$\times$ (resp. 3.3$\times$) faster than Groute (resp. Gunrock), up to 9.5$\times$ (resp. 9.9$\times$).

\eat{
\etitle{CPU performance}.
For all dataset used, \gsf outperforms other CPU-based graph processing systems, and can process large graphs that other systems failed to process.
The reason that \gsf performs better than Gemini, Giraph and GraphX are:
(i) Other systems that employ high-level programming paradigms suffer from abstraction penalties, such as the GAS model, which requires additional copying and caching of messages on the edges. However, \gsf utilizes more fundamental primitives, thereby avoiding such abstraction penalties.
(ii) \gsf uses aggregated communication, which enhances the utilization rate of communication bandwidth while simultaneously reducing random memory access.%\looseness=-1

\etitle{GPU performance}.
For all dataset used, \gsf significantly outperforms other GPU-based graph processing systems using 8 GPUs, especially for the graph traversal applications like BFS.
The reasons that \gsf performs better than Gunrock and Groute are:
(i) they do not take full advantage of NVLink, and their communication implementations are not aware of the asymmetry of the communication topology.
Graph traversal applications like SSSP may suffer severe dynamic load-imbalance problem in various graph datasets.
However, both Gunrock and Groute ignore the NVLink topology and the dynamic load-imbalance, while \gsf enables stealing thus avoiding GPU starvation.
(ii) our system employs adaptive technology, capable of automatically selecting the appropriate inter-GPU and intra-GPU load balancing algorithms based on varying input data and algorithms. This allows for more stable performance.%\looseness=-1
}

\stitle{Exp-4. Graph Learning Performance.}
We evaluate the scalability of the learning stack in ~\gsf~
by training a 3-layer GraphSAGE model on the \texttt{PA} and
\texttt{PD} datasets. The sampling fan-out is set to
$\lbrack 15, 10, 5\rbrack$ and the batch size is $1024$.
\eat{The experimental cluster consists of $4$
nodes, each equipped with a 128-core Intel(R) Xeon(R) Platinum
CPU at 2.90GHz, 768 GB of memory and 4 Nvidia A10 GPUs.}
The nodes allocated in this experiment are
additionally equipped with $4$ Nvidia A10 GPUs.
In both the single-node and distributed experiments, GPU sampling is used
with the number of sampling processes set equal to the number of GPUs.
Figure~\ref{fig:gle-scaleup} shows the results of scaling up
experiments. By increasing the number of GPUs, the end-to-end training time
per epoch decreases linearly. This is because each sampling process handles
a subset of the sampling workloads independently in single-node
training, increasing the number of sampling processes can improve the
training efficiency accordingly. The results of scaling out experiments
are presented in Figure~\ref{fig:gle-scaleout}. The number of used
GPUs in each node is fixed at $2$. By increasing the number of nodes
from $1$ to $4$, we can observe an almost-linear boost in training performance.
This result shows that, despite longer batch execution times in sampling processes
due to network communication costs in distributed sampling and feature collection,
the asynchronous pipelining and prefetch mechanism ensure nearly linear
scaling out performance in the learning stack of ~\gsf~.
  
\subsection{Real-World Applications}

\gsf~has been widely deployed in production at Alibaba,
and it supports about 50,000 graph jobs every day.
Next, we report results from production for the applications
described in Section ~\ref{sec:use-cases}.

\begin{table}[ht]
  \vspace{-1ex}
  \centering
  \caption{Throughput of the real-time fraud detection.}
  \vspace{-2.5ex}
  \label{tab:blacklist}
  \begin{tabular}{|c||c|c|c|c|}
  \hline
  \textbf{\#threads} & \textbf{10} & \textbf{20} & \textbf{30} & \textbf{40} \\
  \hline
  \textbf{Throughput} & 98,907 & 184,826 & 279,005 & 355,813 \\
  \hline
  \end{tabular}
  \vspace*{-1ex}
  \end{table}

  \stitle{Exp-5. Real-time Fraud Detection.}
  Our evaluation of the real-time fraud detection was conducted on an extensive segment of a real-life transaction graph.
  \eat{the scale of which remains confidential due to conflict of interest concerns.}
  Our assessment went beyond the Cypher query discussed in Section~\ref{sec:use-cases}, incorporating
  several queries that explore various relationships.
  \eat{, such as connections among buyers' relatives.
  For this evaluation, we simulated an online deployment scenario, where numerous clients continuously
  placed orders. Each order triggered a set of mandatory queries, reflecting diverse relational checks. }
  In this scenario, numerous clients continuously placed orders, each of which triggered a set of mandatory queries, reflecting diverse relational checks.
  These queries were executed across different CPU thread configurations in \gsf, showcasing a range
  of cloud computational resources from 10 to 40 threads. The results, detailed in Table~\ref{tab:blacklist},
  demonstrate the system's scalable nature, with throughput almost linearly increasing with the number
  of active working threads. This scalability is critical for adapting to varying throughput demands.
  For instance, handling an average load of 10,000 queries per second, and efficiently scaling up to over
  200,000 queries per second during peak hours. Compared to the limitations of previous deployments,
  which struggled under peak load, \gsf~ demonstrates a more than $30\times$ improvement in throughput
  under similar configurations. This highlights \gsf's aptitude for meeting the dynamic and high-volume
  requirements of real-world applications.
  
\stitle{Exp-6. Equity Analysis.}
We evaluated the performance of equity analysis using a graph constructed from open data on the equity relationships of registered companies in China.
This graph includes 0.3 billion vertices and over 1.5 billion edges.
We compared the \gsf~ deployment, as detailed in Section~\ref{sec:use-cases}, against an existing SQL-based baseline.
The baseline, which stored data in relational tables, checked each tuple (\ie a company) and calculated the shares among its shareholders.
Despite incorporating numerous approximations to reduce computational costs,
the baseline was unable to produce complete results.
It allowed only a limited number of tuples to be involved in the queries
and required more than 1 hour to process a small subset of the data.
In contrast, \gsf~ could generate all results within 15 minutes
on the whole graph,
enabling its daily running in production
and ensuring that analysts always have access to the latest results.

\stitle{Exp-7. Social Relation Prediction.}
The performance of prediction was evaluated
by training an NCN model on a in-house social relation dataset,
which consists
of $10$ million vertices and $200$ million edges.
\eat{The training job was conducted using a cluster of $30$ nodes.
each of which has a quad-core CPU and $60$GB of memory.
To optimize the end-to-end throughput,
%  according to the various sampling and training workloads,
 $10$ nodes were allocated for sampling
and the remaining $20$ nodes are used for training. }%eat
The training job was executed on a cluster comprising 30 nodes. In order to optimize end-to-end throughput, tailored to the demands of sampling and training workloads, 10 nodes were designated for sampling while the remaining 20 nodes were utilized for training.
The end-to-end training time per epoch is $1.5$ hours,
which can be linearly scaled for large-scale graphs in production.

\stitle{Exp-8. Cybersecurity Monitoring.}
With \gsf, we upgrade the application reported in~\cite{fan2021graphscope}.
Practice shows that \gsf~ can subsume \gs~ without any performance degradation,
offering a concise and easy-to-maintain deployment instead.
Using graph traversal written in Gremlin,
\gsf~ achieves a speedup of 2,400$\times$ over the equivalent SQL queries\eat{in ODPS}. Since the Trojan detection queries are two-hop graph traversals, they avoid the costly join operations required in SQL queries.

\stitle{Summary.}
We find the following.
(1) The composition and functionality of the components within \gsf~ are both effective and efficient.
(a) \grin~ effectively supports various storage backends with a maximum performance degradation of $8\%$.
(b) \gsf~ outperforms leading systems in LDBC Social Network Benchmarks, achieving $2.45\times$ higher throughput than TuGraph in SNB Interactive workloads,
and an average of $10\times$ smaller latency than TigerGraph in SNB-BI workloads, respectively.
(c) In graph analytics, \gsf~ exceeds both CPU-based (Gemini, Powergraph) and GPU-based (Groute, Gunrock) systems, reaching up to $55.7\times (resp.~9.9\times)$ faster performance.
(d) \gsf~ shows strong scalability in learning tasks, with up to $3.94\times (resp.~3.42\times)$ improvement when scaling from $1(1\times2)$ GPUs to $4(4\times2)$ GPUs.
(2) Better still, \gsf~ performs well in real-world applications.
(a) It manages 355,813 qps in real-time fraud detection, 
and significantly outperforming SQL-based solutions in cybersecurity by over 2,400$\times$.
(b) It facilitates comprehensive full-data analysis in equity analysis. and
(c) It efficiently enables the daily training of social relation prediction models and offers linear scalability.

\section{Conclusion}
\label{conclusion}
\gsf~addresses the limitations of the \emph{``one-size-fits-all''} solution by adopting a modular architecture, and users can selectively deploy components of \gsf~to meet their specific requirements.
% To this end, \gsf~comprises multiple components, and each component is designed to provide specific functionalities.
% Then \gsf~provides a utility tool that enables users to choose specific components, build and generate their respective binaries.
Evaluations on both synthetic and real-life cases 
show that \gsf~can efficiently and flexibly handle diverse application scenarios.
Moving forward, \gsf~ will persist in its evolution towards a high-performance and user-friendly computing system for large-scale graph processing.
%%
%% The next two lines define the bibliography style to be used, and
%% the bibliography file.
\balance
\bibliographystyle{ACM-Reference-Format}
\bibliography{paper}

%%% -*-BibTeX-*-
%%% Do NOT edit. File created by BibTeX with style
%%% ACM-Reference-Format-Journals [18-Jan-2012].

\begin{thebibliography}{92}

%%% ====================================================================
%%% NOTE TO THE USER: you can override these defaults by providing
%%% customized versions of any of these macros before the \bibliography
%%% command.  Each of them MUST provide its own final punctuation,
%%% except for \shownote{}, \showDOI{}, and \showURL{}.  The latter two
%%% do not use final punctuation, in order to avoid confusing it with
%%% the Web address.
%%%
%%% To suppress output of a particular field, define its macro to expand
%%% to an empty string, or better, \unskip, like this:
%%%
%%% \newcommand{\showDOI}[1]{\unskip}   % LaTeX syntax
%%%
%%% \def \showDOI #1{\unskip}           % plain TeX syntax
%%%
%%% ====================================================================

\ifx \showCODEN    \undefined \def \showCODEN     #1{\unskip}     \fi
\ifx \showDOI      \undefined \def \showDOI       #1{#1}\fi
\ifx \showISBNx    \undefined \def \showISBNx     #1{\unskip}     \fi
\ifx \showISBNxiii \undefined \def \showISBNxiii  #1{\unskip}     \fi
\ifx \showISSN     \undefined \def \showISSN      #1{\unskip}     \fi
\ifx \showLCCN     \undefined \def \showLCCN      #1{\unskip}     \fi
\ifx \shownote     \undefined \def \shownote      #1{#1}          \fi
\ifx \showarticletitle \undefined \def \showarticletitle #1{#1}   \fi
\ifx \showURL      \undefined \def \showURL       {\relax}        \fi
% The following commands are used for tagged output and should be
% invisible to TeX
\providecommand\bibfield[2]{#2}
\providecommand\bibinfo[2]{#2}
\providecommand\natexlab[1]{#1}
\providecommand\showeprint[2][]{arXiv:#2}

\bibitem[gir(2011)]%
        {giraph}
 \bibinfo{year}{2011}\natexlab{}.
\newblock \bibinfo{title}{{Apache Giraph}}.
\newblock \bibinfo{howpublished}{\url{https://giraph.apache.org}}.
\newblock


\bibitem[spa(2013)]%
        {sparql}
 \bibinfo{year}{2013}\natexlab{}.
\newblock \bibinfo{title}{W3C Sparql 1.1 Query Language}.
\newblock \bibinfo{howpublished}{\url{https://www.w3.org/TR/sparql11-query/}}.
\newblock


\bibitem[net(2014)]%
        {networkx}
 \bibinfo{year}{2014}\natexlab{}.
\newblock \bibinfo{title}{{NetworkX}}.
\newblock \bibinfo{howpublished}{\url{https://networkx.org/}}.
\newblock


\bibitem[rdf(2014)]%
        {rdf}
 \bibinfo{year}{2014}\natexlab{}.
\newblock \bibinfo{title}{{W3C, Resource Description Framework (RDF)}}.
\newblock \bibinfo{howpublished}{\url{https://www.w3.org/RDF/}}.
\newblock


\bibitem[tin(2015)]%
        {tinkerpop}
 \bibinfo{year}{2015}\natexlab{}.
\newblock \bibinfo{title}{{Apache TinkerPop}}.
\newblock \bibinfo{howpublished}{\url{https://tinkerpop.apache.org}}.
\newblock


\bibitem[cyp(2015)]%
        {cypherneo4j}
 \bibinfo{year}{2015}\natexlab{}.
\newblock \bibinfo{title}{{Cypher Query Language in Neo4j}}.
\newblock
  \bibinfo{howpublished}{\url{https://neo4j.com/product/cypher-graph-query-language/}}.
\newblock


\bibitem[jan(2017)]%
        {janusgraph}
 \bibinfo{year}{2017}\natexlab{}.
\newblock \bibinfo{title}{{JanusGraph}}.
\newblock \bibinfo{howpublished}{\url{https://janusgraph.org/}}.
\newblock


\bibitem[gql(2018)]%
        {gqlstandard}
 \bibinfo{year}{2018}\natexlab{}.
\newblock \bibinfo{title}{{GQL Standard}}.
\newblock \bibinfo{howpublished}{\url{https://https://www.gqlstandards.org}}.
\newblock


\bibitem[gra(2018)]%
        {graphx}
 \bibinfo{year}{2018}\natexlab{}.
\newblock \bibinfo{title}{{Spark GraphX}}.
\newblock \bibinfo{howpublished}{\url{https://spark.apache.org/graphx/}}.
\newblock


\bibitem[ldb(2019)]%
        {ldbc-snb}
 \bibinfo{year}{2019}\natexlab{}.
\newblock \bibinfo{title}{{LDBC Social Network Benchmark}}.
\newblock \bibinfo{howpublished}{\url{https://ldbcouncil.org/benchmarks/snb/}}.
\newblock


\bibitem[tig(2019)]%
        {tigergraph}
 \bibinfo{year}{2019}\natexlab{}.
\newblock \bibinfo{title}{{TigerGraph}}.
\newblock \bibinfo{howpublished}{\url{https://www.tigergraph.com/}}.
\newblock


\bibitem[gs-(2022)]%
        {gs-ldbc-analysis-report}
 \bibinfo{year}{2022}\natexlab{}.
\newblock \bibinfo{title}{Performance report of LDBC Graphalytics}.
\newblock
  \bibinfo{howpublished}{\url{https://github.com/alibaba/libgrape-lite/blob/master/Performance.md}}.
\newblock


\bibitem[ORC(2023)]%
        {ORC}
 \bibinfo{year}{2023}\natexlab{}.
\newblock \bibinfo{title}{{Apache ORC}}.
\newblock \bibinfo{howpublished}{\url{https://orc.apache.org}}.
\newblock


\bibitem[Par(2023)]%
        {Parquet}
 \bibinfo{year}{2023}\natexlab{}.
\newblock \bibinfo{title}{{Apache Parquet}}.
\newblock \bibinfo{howpublished}{\url{https://parquet.apache.org/}}.
\newblock


\bibitem[ldb(2023a)]%
        {ldbc-ia-audit}
 \bibinfo{year}{2023}\natexlab{a}.
\newblock \bibinfo{title}{{Auditing results of LDBC SNB Interactive Workload}}.
\newblock
  \bibinfo{howpublished}{\url{https://ldbcouncil.org/benchmarks/snb-interactive/}}.
\newblock


\bibitem[tig(2023)]%
        {tiger-audit}
 \bibinfo{year}{2023}\natexlab{}.
\newblock \bibinfo{title}{{Full Disclosure Report for TigerGraph of the LDBC
  Social Network Benchmark}}.
\newblock
  \bibinfo{howpublished}{\url{https://ldbcouncil.org/benchmarks/snb/LDBC_SNB_BI_20221109_SF1000_tigergraph.pdf}}.
\newblock


\bibitem[Gra(2023)]%
        {GraphAr}
 \bibinfo{year}{2023}\natexlab{}.
\newblock \bibinfo{title}{{GraphAr}}.
\newblock \bibinfo{howpublished}{\url{https://github.com/alibaba/GraphAr}}.
\newblock


\bibitem[gql(2023)]%
        {gql}
 \bibinfo{year}{2023}\natexlab{}.
\newblock \bibinfo{title}{ISO graph query standard GQL}.
\newblock \bibinfo{howpublished}{\url{https://www.gqlstandards.org/}}.
\newblock


\bibitem[ldb(2023b)]%
        {ldbc-analytics}
 \bibinfo{year}{2023}\natexlab{b}.
\newblock \bibinfo{title}{{LDBC Graphalytics}}.
\newblock
  \bibinfo{howpublished}{\url{https://ldbcouncil.org/benchmarks/graphalytics/}}.
\newblock


\bibitem[neo(2023)]%
        {neo4j}
 \bibinfo{year}{2023}\natexlab{}.
\newblock \bibinfo{title}{{Neo4j}}.
\newblock \bibinfo{howpublished}{\url{https://neo4j.com/}}.
\newblock


\bibitem[ont(2023)]%
        {ontotext}
 \bibinfo{year}{2023}\natexlab{}.
\newblock \bibinfo{title}{OntoText}.
\newblock \bibinfo{howpublished}{\url{https://www.ontotext.com/}}.
\newblock


\bibitem[tor(2023)]%
        {torch}
 \bibinfo{year}{2023}\natexlab{}.
\newblock \bibinfo{title}{{PyTorch}}.
\newblock \bibinfo{howpublished}{\url{https://github.com/pytorch/pytorch}}.
\newblock


\bibitem[pyg(2023)]%
        {pyg}
 \bibinfo{year}{2023}\natexlab{}.
\newblock \bibinfo{title}{{PyTorch Geometric}}.
\newblock
  \bibinfo{howpublished}{\url{https://github.com/pyg-team/pytorch_geometric}}.
\newblock


\bibitem[opt(2023)]%
        {opt-queries}
 \bibinfo{year}{2023}\natexlab{}.
\newblock \bibinfo{title}{Queries used for the experiment of graph query
  optimization}.
\newblock
  \bibinfo{howpublished}{\url{https://github.com/alibaba/GraphScope/tree/main/flex/resources/queries/examples/store_procedure}}.
\newblock


\bibitem[tf(2023)]%
        {tf}
 \bibinfo{year}{2023}\natexlab{}.
\newblock \bibinfo{title}{{TensorFlow}}.
\newblock
  \bibinfo{howpublished}{\url{https://github.com/tensorflow/tensorflow}}.
\newblock


\bibitem[ogb(2023)]%
        {ogbn}
 \bibinfo{year}{2023}\natexlab{}.
\newblock \bibinfo{title}{{The Open Graph Benchmark (OGB)}}.
\newblock \bibinfo{howpublished}{\url{https://ogb.stanford.edu/}}.
\newblock


\bibitem[tug(2023)]%
        {tugraph}
 \bibinfo{year}{2023}\natexlab{}.
\newblock \bibinfo{title}{{TuGraph, The Distributed Graph Database Behind
  AliPay.}}
\newblock \bibinfo{howpublished}{\url{https://tugraph.antgroup.com/}}.
\newblock


\bibitem[Angles et~al\mbox{.}(2017)]%
        {graphquerysurvey}
\bibfield{author}{\bibinfo{person}{Renzo Angles}, \bibinfo{person}{Marcelo
  Arenas}, \bibinfo{person}{Pablo Barcel\'{o}}, \bibinfo{person}{Aidan Hogan},
  \bibinfo{person}{Juan Reutter}, {and} \bibinfo{person}{Domagoj Vrgo\v{c}}.}
  \bibinfo{year}{2017}\natexlab{}.
\newblock \showarticletitle{Foundations of Modern Query Languages for Graph
  Databases}.
\newblock \bibinfo{journal}{\emph{ACM Comput. Surv.}} \bibinfo{volume}{50},
  \bibinfo{number}{5}, Article \bibinfo{articleno}{68} (\bibinfo{date}{sep}
  \bibinfo{year}{2017}), \bibinfo{numpages}{40}~pages.
\newblock


\bibitem[Angles et~al\mbox{.}(2023)]%
        {angles2023pg}
\bibfield{author}{\bibinfo{person}{Renzo Angles}, \bibinfo{person}{Angela
  Bonifati}, \bibinfo{person}{Stefania Dumbrava}, \bibinfo{person}{George
  Fletcher}, \bibinfo{person}{Alastair Green}, \bibinfo{person}{Jan Hidders},
  \bibinfo{person}{Bei Li}, \bibinfo{person}{Leonid Libkin},
  \bibinfo{person}{Victor Marsault}, \bibinfo{person}{Wim Martens},
  {et~al\mbox{.}}} \bibinfo{year}{2023}\natexlab{}.
\newblock \showarticletitle{PG-Schema: Schemas for property graphs}.
\newblock \bibinfo{journal}{\emph{Proceedings of the ACM on Management of
  Data}} \bibinfo{volume}{1}, \bibinfo{number}{2} (\bibinfo{year}{2023}),
  \bibinfo{pages}{1--25}.
\newblock


\bibitem[Anikin et~al\mbox{.}(2019)]%
        {anikin2019labeled}
\bibfield{author}{\bibinfo{person}{Dmitry Anikin}, \bibinfo{person}{Oleg
  Borisenko}, {and} \bibinfo{person}{Yaroslav Nedumov}.}
  \bibinfo{year}{2019}\natexlab{}.
\newblock \showarticletitle{Labeled property graphs: SQL or NoSQL?}. In
  \bibinfo{booktitle}{\emph{2019 Ivannikov Memorial Workshop (IVMEM)}}. IEEE,
  \bibinfo{pages}{7--13}.
\newblock


\bibitem[Azadifar et~al\mbox{.}(2022)]%
        {azadifar2022graph}
\bibfield{author}{\bibinfo{person}{Saeid Azadifar}, \bibinfo{person}{Mehrdad
  Rostami}, \bibinfo{person}{Kamal Berahmand}, \bibinfo{person}{Parham Moradi},
  {and} \bibinfo{person}{Mourad Oussalah}.} \bibinfo{year}{2022}\natexlab{}.
\newblock \showarticletitle{Graph-based relevancy-redundancy gene selection
  method for cancer diagnosis}.
\newblock \bibinfo{journal}{\emph{Computers in Biology and Medicine}}
  \bibinfo{volume}{147} (\bibinfo{year}{2022}), \bibinfo{pages}{105766}.
\newblock


\bibitem[Baken(2020)]%
        {baken2020linked}
\bibfield{author}{\bibinfo{person}{Nico Baken}.}
  \bibinfo{year}{2020}\natexlab{}.
\newblock \showarticletitle{Linked data for smart homes: Comparing RDF and
  labeled property graphs}. In \bibinfo{booktitle}{\emph{LDAC2020—8th Linked
  Data in Architecture and Construction Workshop}}. \bibinfo{pages}{23--36}.
\newblock


\bibitem[Ben-Nun et~al\mbox{.}(2017)]%
        {Ben2017Groute}
\bibfield{author}{\bibinfo{person}{Tal Ben-Nun}, \bibinfo{person}{Michael
  Sutton}, \bibinfo{person}{Sreepathi Pai}, {and} \bibinfo{person}{Keshav
  Pingali}.} \bibinfo{year}{2017}\natexlab{}.
\newblock \showarticletitle{Groute: An Asynchronous Multi-GPU Programming Model
  for Irregular Computations}. In \bibinfo{booktitle}{\emph{ACM Sigplan
  Symposium on Principles and Practice of Parallel Programming}}.
  \bibinfo{pages}{235--248}.
\newblock


\bibitem[Blondel et~al\mbox{.}(2008)]%
        {blondel2008fast}
\bibfield{author}{\bibinfo{person}{Vincent~D Blondel},
  \bibinfo{person}{Jean-Loup Guillaume}, \bibinfo{person}{Renaud Lambiotte},
  {and} \bibinfo{person}{Etienne Lefebvre}.} \bibinfo{year}{2008}\natexlab{}.
\newblock \showarticletitle{Fast unfolding of communities in large networks}.
\newblock \bibinfo{journal}{\emph{Journal of statistical mechanics: theory and
  experiment}} \bibinfo{volume}{2008}, \bibinfo{number}{10}
  (\bibinfo{year}{2008}), \bibinfo{pages}{P10008}.
\newblock


\bibitem[Chen et~al\mbox{.}(2019)]%
        {chen2019powerlyra}
\bibfield{author}{\bibinfo{person}{Rong Chen}, \bibinfo{person}{Jiaxin Shi},
  \bibinfo{person}{Yanzhe Chen}, \bibinfo{person}{Binyu Zang},
  \bibinfo{person}{Haibing Guan}, {and} \bibinfo{person}{Haibo Chen}.}
  \bibinfo{year}{2019}\natexlab{}.
\newblock \showarticletitle{Powerlyra: Differentiated graph computation and
  partitioning on skewed graphs}.
\newblock \bibinfo{journal}{\emph{ACM Transactions on Parallel Computing
  (TOPC)}} \bibinfo{volume}{5}, \bibinfo{number}{3} (\bibinfo{year}{2019}),
  \bibinfo{pages}{1--39}.
\newblock


\bibitem[Davis(2019)]%
        {graphblas}
\bibfield{author}{\bibinfo{person}{Timothy~A. Davis}.}
  \bibinfo{year}{2019}\natexlab{}.
\newblock \showarticletitle{Algorithm 1000: SuiteSparse:GraphBLAS: Graph
  Algorithms in the Language of Sparse Linear Algebra}.
\newblock \bibinfo{journal}{\emph{ACM Trans. Math. Softw.}}
  \bibinfo{volume}{45}, \bibinfo{number}{4} (\bibinfo{year}{2019}).
\newblock


\bibitem[Davis and Hu(2011)]%
        {florida-spmat-datasets}
\bibfield{author}{\bibinfo{person}{Timothy~A. Davis} {and}
  \bibinfo{person}{Yifan Hu}.} \bibinfo{year}{2011}\natexlab{}.
\newblock \showarticletitle{The University of Florida Sparse Matrix
  Collection}.
\newblock \bibinfo{journal}{\emph{ACM Trans. Math. Softw.}}
  \bibinfo{volume}{38}, \bibinfo{number}{1}, Article \bibinfo{articleno}{1}
  (\bibinfo{date}{dec} \bibinfo{year}{2011}), \bibinfo{numpages}{25}~pages.
\newblock
\showISSN{0098-3500}
\urldef\tempurl%
\url{https://doi.org/10.1145/2049662.2049663}
\showDOI{\tempurl}


\bibitem[De~Leo and Boncz(2021)]%
        {de2021teseo}
\bibfield{author}{\bibinfo{person}{Dean De~Leo} {and} \bibinfo{person}{Peter
  Boncz}.} \bibinfo{year}{2021}\natexlab{}.
\newblock \showarticletitle{Teseo and the analysis of structural dynamic
  graphs}.
\newblock \bibinfo{journal}{\emph{Proceedings of the VLDB Endowment}}
  \bibinfo{volume}{14}, \bibinfo{number}{6} (\bibinfo{year}{2021}),
  \bibinfo{pages}{1053--1066}.
\newblock


\bibitem[Decker et~al\mbox{.}(2000)]%
        {decker2000semantic}
\bibfield{author}{\bibinfo{person}{Stefan Decker}, \bibinfo{person}{Sergey
  Melnik}, \bibinfo{person}{Frank Van~Harmelen}, \bibinfo{person}{Dieter
  Fensel}, \bibinfo{person}{Michel Klein}, \bibinfo{person}{Jeen Broekstra},
  \bibinfo{person}{Michael Erdmann}, {and} \bibinfo{person}{Ian Horrocks}.}
  \bibinfo{year}{2000}\natexlab{}.
\newblock \showarticletitle{The semantic web: The roles of XML and RDF}.
\newblock \bibinfo{journal}{\emph{IEEE Internet computing}}
  \bibinfo{volume}{4}, \bibinfo{number}{5} (\bibinfo{year}{2000}),
  \bibinfo{pages}{63--73}.
\newblock


\bibitem[Dhulipala et~al\mbox{.}(2019)]%
        {dhulipala2019low}
\bibfield{author}{\bibinfo{person}{Laxman Dhulipala}, \bibinfo{person}{Guy~E
  Blelloch}, {and} \bibinfo{person}{Julian Shun}.}
  \bibinfo{year}{2019}\natexlab{}.
\newblock \showarticletitle{Low-latency graph streaming using compressed
  purely-functional trees}. In \bibinfo{booktitle}{\emph{Proceedings of the
  40th ACM SIGPLAN conference on programming language design and
  implementation}}. \bibinfo{pages}{918--934}.
\newblock


\bibitem[Ediger et~al\mbox{.}(2012)]%
        {ediger2012stinger}
\bibfield{author}{\bibinfo{person}{David Ediger}, \bibinfo{person}{Rob McColl},
  \bibinfo{person}{Jason Riedy}, {and} \bibinfo{person}{David~A Bader}.}
  \bibinfo{year}{2012}\natexlab{}.
\newblock \showarticletitle{Stinger: High performance data structure for
  streaming graphs}. In \bibinfo{booktitle}{\emph{2012 IEEE Conference on High
  Performance Extreme Computing}}. IEEE, \bibinfo{pages}{1--5}.
\newblock


\bibitem[Erling et~al\mbox{.}(2015)]%
        {snb-interactive}
\bibfield{author}{\bibinfo{person}{Orri Erling}, \bibinfo{person}{Alex
  Averbuch}, \bibinfo{person}{Josep Larriba-Pey}, \bibinfo{person}{Hassan
  Chafi}, \bibinfo{person}{Andrey Gubichev}, \bibinfo{person}{Arnau Prat},
  \bibinfo{person}{Minh-Duc Pham}, {and} \bibinfo{person}{Peter Boncz}.}
  \bibinfo{year}{2015}\natexlab{}.
\newblock \showarticletitle{The LDBC Social Network Benchmark: Interactive
  Workload}. In \bibinfo{booktitle}{\emph{Proceedings of the 2015 ACM SIGMOD
  International Conference on Management of Data}}
  \emph{(\bibinfo{series}{SIGMOD '15})}. \bibinfo{publisher}{Association for
  Computing Machinery}, \bibinfo{address}{New York, NY, USA},
  \bibinfo{pages}{619–630}.
\newblock


\bibitem[Fan et~al\mbox{.}(2021)]%
        {fan2021graphscope}
\bibfield{author}{\bibinfo{person}{Wenfei Fan}, \bibinfo{person}{Tao He},
  \bibinfo{person}{Longbin Lai}, \bibinfo{person}{Xue Li},
  \bibinfo{person}{Yong Li}, \bibinfo{person}{Zhao Li},
  \bibinfo{person}{Zhengping Qian}, \bibinfo{person}{Chao Tian},
  \bibinfo{person}{Lei Wang}, \bibinfo{person}{Jingbo Xu}, {et~al\mbox{.}}}
  \bibinfo{year}{2021}\natexlab{}.
\newblock \showarticletitle{GraphScope: a unified engine for big graph
  processing}.
\newblock \bibinfo{journal}{\emph{Proceedings of the VLDB Endowment}}
  \bibinfo{volume}{14}, \bibinfo{number}{12} (\bibinfo{year}{2021}),
  \bibinfo{pages}{2879--2892}.
\newblock


\bibitem[Fan et~al\mbox{.}(2018)]%
        {fan2018parallelizing}
\bibfield{author}{\bibinfo{person}{Wenfei Fan}, \bibinfo{person}{Wenyuan Yu},
  \bibinfo{person}{Jingbo Xu}, \bibinfo{person}{Jingren Zhou},
  \bibinfo{person}{Xiaojian Luo}, \bibinfo{person}{Qiang Yin},
  \bibinfo{person}{Ping Lu}, \bibinfo{person}{Yang Cao}, {and}
  \bibinfo{person}{Ruiqi Xu}.} \bibinfo{year}{2018}\natexlab{}.
\newblock \showarticletitle{Parallelizing sequential graph computations}.
\newblock \bibinfo{journal}{\emph{ACM Transactions on Database Systems (TODS)}}
  \bibinfo{volume}{43}, \bibinfo{number}{4} (\bibinfo{year}{2018}),
  \bibinfo{pages}{1--39}.
\newblock


\bibitem[Francis et~al\mbox{.}(2018)]%
        {francis2018cypher}
\bibfield{author}{\bibinfo{person}{Nadime Francis}, \bibinfo{person}{Alastair
  Green}, \bibinfo{person}{Paolo Guagliardo}, \bibinfo{person}{Leonid Libkin},
  \bibinfo{person}{Tobias Lindaaker}, \bibinfo{person}{Victor Marsault},
  \bibinfo{person}{Stefan Plantikow}, \bibinfo{person}{Mats Rydberg},
  \bibinfo{person}{Petra Selmer}, {and} \bibinfo{person}{Andr{\'e}s Taylor}.}
  \bibinfo{year}{2018}\natexlab{}.
\newblock \showarticletitle{Cypher: An evolving query language for property
  graphs}. In \bibinfo{booktitle}{\emph{Proceedings of the SIGMOD 2018}}.
  \bibinfo{pages}{1433--1445}.
\newblock


\bibitem[Fuchs et~al\mbox{.}(2022)]%
        {Sortledton}
\bibfield{author}{\bibinfo{person}{Per Fuchs}, \bibinfo{person}{Domagoj
  Margan}, {and} \bibinfo{person}{Jana Giceva}.}
  \bibinfo{year}{2022}\natexlab{}.
\newblock \showarticletitle{Sortledton: A Universal, Transactional Graph Data
  Structure}.
\newblock \bibinfo{journal}{\emph{Proc. VLDB Endow.}} \bibinfo{volume}{15},
  \bibinfo{number}{6} (\bibinfo{year}{2022}), \bibinfo{pages}{1173–1186}.
\newblock


\bibitem[Gandhi and Iyer(2021)]%
        {gandhi2021p3}
\bibfield{author}{\bibinfo{person}{Swapnil Gandhi} {and}
  \bibinfo{person}{Anand~Padmanabha Iyer}.} \bibinfo{year}{2021}\natexlab{}.
\newblock \showarticletitle{P3: Distributed deep graph learning at scale}. In
  \bibinfo{booktitle}{\emph{15th $\{$USENIX$\}$ OSDI 21}}.
  \bibinfo{pages}{551--568}.
\newblock


\bibitem[Gong et~al\mbox{.}(2021)]%
        {ingress}
\bibfield{author}{\bibinfo{person}{Shufeng Gong}, \bibinfo{person}{Chao Tian},
  \bibinfo{person}{Qiang Yin}, \bibinfo{person}{Wenyuan Yu},
  \bibinfo{person}{Yanfeng Zhang}, \bibinfo{person}{Liang Geng},
  \bibinfo{person}{Song Yu}, \bibinfo{person}{Ge Yu}, {and}
  \bibinfo{person}{Jingren Zhou}.} \bibinfo{year}{2021}\natexlab{}.
\newblock \showarticletitle{Automating Incremental Graph Processing with
  Flexible Memoization}.
\newblock \bibinfo{journal}{\emph{Proc. VLDB Endow.}} \bibinfo{volume}{14},
  \bibinfo{number}{9} (\bibinfo{date}{may} \bibinfo{year}{2021}),
  \bibinfo{pages}{1613–1625}.
\newblock
\showISSN{2150-8097}


\bibitem[Gonzalez et~al\mbox{.}(2012)]%
        {gonzalez2012powergraph}
\bibfield{author}{\bibinfo{person}{Joseph~E Gonzalez}, \bibinfo{person}{Yucheng
  Low}, \bibinfo{person}{Haijie Gu}, \bibinfo{person}{Danny Bickson}, {and}
  \bibinfo{person}{Carlos Guestrin}.} \bibinfo{year}{2012}\natexlab{}.
\newblock \showarticletitle{PowerGraph: Distributed Graph-Parallel Computation
  on Natural Graphs.}. In \bibinfo{booktitle}{\emph{OSDI}},
  Vol.~\bibinfo{volume}{12}. \bibinfo{pages}{2}.
\newblock


\bibitem[Guo and Wang(2020)]%
        {guo2020deep}
\bibfield{author}{\bibinfo{person}{Zhiwei Guo} {and} \bibinfo{person}{Heng
  Wang}.} \bibinfo{year}{2020}\natexlab{}.
\newblock \showarticletitle{A deep graph neural network-based mechanism for
  social recommendations}.
\newblock \bibinfo{journal}{\emph{IEEE Transactions on Industrial Informatics}}
  \bibinfo{volume}{17}, \bibinfo{number}{4} (\bibinfo{year}{2020}),
  \bibinfo{pages}{2776--2783}.
\newblock


\bibitem[Holme(2017)]%
        {holme2017three}
\bibfield{author}{\bibinfo{person}{Petter Holme}.}
  \bibinfo{year}{2017}\natexlab{}.
\newblock \showarticletitle{Three faces of node importance in network
  epidemiology: Exact results for small graphs}.
\newblock \bibinfo{journal}{\emph{Physical Review E}} \bibinfo{volume}{96},
  \bibinfo{number}{6} (\bibinfo{year}{2017}), \bibinfo{pages}{062305}.
\newblock


\bibitem[Jamshidi et~al\mbox{.}(2020)]%
        {jamshidi2020peregrine}
\bibfield{author}{\bibinfo{person}{Kasra Jamshidi}, \bibinfo{person}{Rakesh
  Mahadasa}, {and} \bibinfo{person}{Keval Vora}.}
  \bibinfo{year}{2020}\natexlab{}.
\newblock \showarticletitle{Peregrine: a pattern-aware graph mining system}. In
  \bibinfo{booktitle}{\emph{Proceedings of the Fifteenth European Conference on
  Computer Systems}}. \bibinfo{pages}{1--16}.
\newblock


\bibitem[Kumar and Huang(2020)]%
        {kumar2020graphone}
\bibfield{author}{\bibinfo{person}{Pradeep Kumar} {and}
  \bibinfo{person}{H~Howie Huang}.} \bibinfo{year}{2020}\natexlab{}.
\newblock \showarticletitle{Graphone: A data store for real-time analytics on
  evolving graphs}.
\newblock \bibinfo{journal}{\emph{ACM Transactions on Storage (TOS)}}
  \bibinfo{volume}{15}, \bibinfo{number}{4} (\bibinfo{year}{2020}),
  \bibinfo{pages}{1--40}.
\newblock


\bibitem[Lai et~al\mbox{.}(2023)]%
        {glogs}
\bibfield{author}{\bibinfo{person}{Longbin Lai}, \bibinfo{person}{Yufan Yang},
  \bibinfo{person}{Zhibin Wang}, \bibinfo{person}{Yuxuan Liu},
  \bibinfo{person}{Haotian Ma}, \bibinfo{person}{Sijie Shen},
  \bibinfo{person}{Bingqing Lyu}, \bibinfo{person}{Xiaoli Zhou},
  \bibinfo{person}{Wenyuan Yu}, \bibinfo{person}{Zhengping Qian},
  \bibinfo{person}{Chen Tian}, \bibinfo{person}{Sheng Zhong},
  \bibinfo{person}{Yeh-Ching Chung}, {and} \bibinfo{person}{Jingren Zhou}.}
  \bibinfo{year}{2023}\natexlab{}.
\newblock \showarticletitle{{GLogS}: Interactive Graph Pattern Matching Query
  At Large Scale}. In \bibinfo{booktitle}{\emph{2023 USENIX Annual Technical
  Conference (USENIX ATC 23)}}. \bibinfo{publisher}{USENIX Association},
  \bibinfo{address}{Boston, MA}, \bibinfo{pages}{53--69}.
\newblock


\bibitem[Leskovec and Krevl(2014)]%
        {snapnets}
\bibfield{author}{\bibinfo{person}{Jure Leskovec} {and} \bibinfo{person}{Andrej
  Krevl}.} \bibinfo{year}{2014}\natexlab{}.
\newblock \bibinfo{title}{{SNAP Datasets}: {Stanford} Large Network Dataset
  Collection}.
\newblock \bibinfo{howpublished}{\url{http://snap.stanford.edu/data}}.
\newblock


\bibitem[Li et~al\mbox{.}(2022)]%
        {li2022graph}
\bibfield{author}{\bibinfo{person}{Michelle~M Li}, \bibinfo{person}{Kexin
  Huang}, {and} \bibinfo{person}{Marinka Zitnik}.}
  \bibinfo{year}{2022}\natexlab{}.
\newblock \showarticletitle{Graph representation learning in biomedicine and
  healthcare}.
\newblock \bibinfo{journal}{\emph{Nature Biomedical Engineering}}
  \bibinfo{volume}{6}, \bibinfo{number}{12} (\bibinfo{year}{2022}),
  \bibinfo{pages}{1353--1369}.
\newblock


\bibitem[Li et~al\mbox{.}(2023a)]%
        {hiactor}
\bibfield{author}{\bibinfo{person}{Su Li} {et~al\mbox{.}}}
  \bibinfo{year}{2023}\natexlab{a}.
\newblock \bibinfo{title}{Hiactor: an open-source hierarchical actor
  framework}.
\newblock \bibinfo{howpublished}{\url{https://github.com/alibaba/hiactor}}.
\newblock


\bibitem[Li et~al\mbox{.}(2023b)]%
        {li2023flash}
\bibfield{author}{\bibinfo{person}{Xue Li}, \bibinfo{person}{Ke Meng},
  \bibinfo{person}{Lu Qin}, \bibinfo{person}{Longbin Lai},
  \bibinfo{person}{Wenyuan Yu}, \bibinfo{person}{Zhengping Qian},
  \bibinfo{person}{Xuemin Lin}, {and} \bibinfo{person}{Jingren Zhou}.}
  \bibinfo{year}{2023}\natexlab{b}.
\newblock \showarticletitle{FLASH: A Framework for Programming Distributed
  Graph Processing Algorithms}. In \bibinfo{booktitle}{\emph{2023 IEEE 39th
  International Conference on Data Engineering (ICDE)}}. IEEE,
  \bibinfo{pages}{232--244}.
\newblock


\bibitem[Liu et~al\mbox{.}(2023)]%
        {liu2023bgl}
\bibfield{author}{\bibinfo{person}{Tianfeng Liu}, \bibinfo{person}{Yangrui
  Chen}, \bibinfo{person}{Dan Li}, \bibinfo{person}{Chuan Wu},
  \bibinfo{person}{Yibo Zhu}, \bibinfo{person}{Jun He},
  \bibinfo{person}{Yanghua Peng}, \bibinfo{person}{Hongzheng Chen},
  \bibinfo{person}{Hongzhi Chen}, {and} \bibinfo{person}{Chuanxiong Guo}.}
  \bibinfo{year}{2023}\natexlab{}.
\newblock \showarticletitle{$\{$BGL$\}$:$\{$GPU-Efficient$\}$$\{$GNN$\}$
  Training by Optimizing Graph Data $\{$I/O$\}$ and Preprocessing}. In
  \bibinfo{booktitle}{\emph{20th USENIX Symposium on NSDI 23}}.
  \bibinfo{pages}{103--118}.
\newblock


\bibitem[Liu et~al\mbox{.}(2022)]%
        {liu2022federated}
\bibfield{author}{\bibinfo{person}{Zhiwei Liu}, \bibinfo{person}{Liangwei
  Yang}, \bibinfo{person}{Ziwei Fan}, \bibinfo{person}{Hao Peng}, {and}
  \bibinfo{person}{Philip~S Yu}.} \bibinfo{year}{2022}\natexlab{}.
\newblock \showarticletitle{Federated social recommendation with graph neural
  network}.
\newblock \bibinfo{journal}{\emph{ACM Transactions on Intelligent Systems and
  Technology (TIST)}} \bibinfo{volume}{13}, \bibinfo{number}{4}
  (\bibinfo{year}{2022}), \bibinfo{pages}{1--24}.
\newblock


\bibitem[Macko et~al\mbox{.}(2015)]%
        {macko2015llama}
\bibfield{author}{\bibinfo{person}{Peter Macko}, \bibinfo{person}{Virendra~J
  Marathe}, \bibinfo{person}{Daniel~W Margo}, {and} \bibinfo{person}{Margo~I
  Seltzer}.} \bibinfo{year}{2015}\natexlab{}.
\newblock \showarticletitle{Llama: Efficient graph analytics using large
  multiversioned arrays}. In \bibinfo{booktitle}{\emph{2015 IEEE 31st
  International Conference on Data Engineering}}. IEEE,
  \bibinfo{pages}{363--374}.
\newblock


\bibitem[Malewicz et~al\mbox{.}(2010)]%
        {malewicz2010pregel}
\bibfield{author}{\bibinfo{person}{Grzegorz Malewicz},
  \bibinfo{person}{Matthew~H Austern}, \bibinfo{person}{Aart~JC Bik},
  \bibinfo{person}{James~C Dehnert}, \bibinfo{person}{Ilan Horn},
  \bibinfo{person}{Naty Leiser}, {and} \bibinfo{person}{Grzegorz Czajkowski}.}
  \bibinfo{year}{2010}\natexlab{}.
\newblock \showarticletitle{Pregel: a system for large-scale graph processing}.
  In \bibinfo{booktitle}{\emph{Proceedings of the 2010 ACM SIGMOD International
  Conference on Management of data}}. \bibinfo{pages}{135--146}.
\newblock


\bibitem[McCune et~al\mbox{.}(2015)]%
        {mccune2015thinking}
\bibfield{author}{\bibinfo{person}{Robert~Ryan McCune}, \bibinfo{person}{Tim
  Weninger}, {and} \bibinfo{person}{Greg Madey}.}
  \bibinfo{year}{2015}\natexlab{}.
\newblock \showarticletitle{Thinking like a vertex: a survey of vertex-centric
  frameworks for large-scale distributed graph processing}.
\newblock \bibinfo{journal}{\emph{ACM Computing Surveys (CSUR)}}
  \bibinfo{volume}{48}, \bibinfo{number}{2} (\bibinfo{year}{2015}),
  \bibinfo{pages}{1--39}.
\newblock


\bibitem[Meng et~al\mbox{.}(2023)]%
        {gum}
\bibfield{author}{\bibinfo{person}{Ke Meng}, \bibinfo{person}{Liang Geng},
  \bibinfo{person}{Xue Li}, \bibinfo{person}{Qian Tao},
  \bibinfo{person}{Wenyuan Yu}, {and} \bibinfo{person}{Jingren Zhou}.}
  \bibinfo{year}{2023}\natexlab{}.
\newblock \showarticletitle{Efficient Multi-GPU Graph Processing with Remote
  Work Stealing}. In \bibinfo{booktitle}{\emph{2023 IEEE 39th International
  Conference on Data Engineering (ICDE)}}. \bibinfo{pages}{191--204}.
\newblock


\bibitem[Meng et~al\mbox{.}(2019)]%
        {Meng2019Gswitch}
\bibfield{author}{\bibinfo{person}{Ke Meng}, \bibinfo{person}{Jiajia Li},
  \bibinfo{person}{Guangming Tan}, {and} \bibinfo{person}{Ninghui Sun}.}
  \bibinfo{year}{2019}\natexlab{}.
\newblock \showarticletitle{A Pattern Based Algorithmic Autotuner for Graph
  Processing on GPUs}. In \bibinfo{booktitle}{\emph{Proceedings of the 24th
  Symposium on Principles and Practice of Parallel Programming}} (Washington,
  District of Columbia) \emph{(\bibinfo{series}{PPoPP '19})}.
  \bibinfo{publisher}{ACM}, \bibinfo{address}{New York, NY, USA},
  \bibinfo{pages}{201–213}.
\newblock
\showISBNx{9781450362252}


\bibitem[Page et~al\mbox{.}(1998)]%
        {page1998pagerank}
\bibfield{author}{\bibinfo{person}{Lawrence Page}, \bibinfo{person}{Sergey
  Brin}, \bibinfo{person}{Rajeev Motwani}, {and} \bibinfo{person}{Terry
  Winograd}.} \bibinfo{year}{1998}\natexlab{}.
\newblock \bibinfo{booktitle}{\emph{The pagerank citation ranking: Bring order
  to the web}}.
\newblock \bibinfo{type}{{T}echnical {R}eport}. \bibinfo{institution}{Technical
  report, stanford University}.
\newblock


\bibitem[Pandey et~al\mbox{.}(2021)]%
        {terrace}
\bibfield{author}{\bibinfo{person}{Prashant Pandey}, \bibinfo{person}{Brian
  Wheatman}, \bibinfo{person}{Helen Xu}, {and} \bibinfo{person}{Aydin Buluc}.}
  \bibinfo{year}{2021}\natexlab{}.
\newblock \showarticletitle{Terrace: A Hierarchical Graph Container for Skewed
  Dynamic Graphs}. In \bibinfo{booktitle}{\emph{Proceedings of the 2021
  SIGMOD}} (Virtual Event, China) \emph{(\bibinfo{series}{SIGMOD '21})}.
  \bibinfo{pages}{1372–1385}.
\newblock


\bibitem[Pandey et~al\mbox{.}(2020)]%
        {pandey2020c}
\bibfield{author}{\bibinfo{person}{Santosh Pandey}, \bibinfo{person}{Lingda
  Li}, \bibinfo{person}{Adolfy Hoisie}, \bibinfo{person}{Xiaoye~S Li}, {and}
  \bibinfo{person}{Hang Liu}.} \bibinfo{year}{2020}\natexlab{}.
\newblock \showarticletitle{C-SAW: A framework for graph sampling and random
  walk on GPUs}. In \bibinfo{booktitle}{\emph{SC20: International Conference
  for High Performance Computing, Networking, Storage and Analysis}}. IEEE,
  \bibinfo{pages}{1--15}.
\newblock


\bibitem[Qian et~al\mbox{.}(2021)]%
        {qian2021gaia}
\bibfield{author}{\bibinfo{person}{Zhengping Qian}, \bibinfo{person}{Chenqiang
  Min}, \bibinfo{person}{Longbin Lai}, \bibinfo{person}{Yong Fang},
  \bibinfo{person}{Gaofeng Li}, \bibinfo{person}{Youyang Yao},
  \bibinfo{person}{Bingqing Lyu}, \bibinfo{person}{Xiaoli Zhou},
  \bibinfo{person}{Zhimin Chen}, {and} \bibinfo{person}{Jingren Zhou}.}
  \bibinfo{year}{2021}\natexlab{}.
\newblock \showarticletitle{{GAIA}: A System for Interactive Analysis on
  Distributed Graphs Using a {High-Level} Language}. In
  \bibinfo{booktitle}{\emph{18th USENIX Symposium on Networked Systems Design
  and Implementation (NSDI 21)}}. \bibinfo{publisher}{USENIX Association},
  \bibinfo{pages}{321--335}.
\newblock
\showISBNx{978-1-939133-21-2}


\bibitem[Rodriguez(2015)]%
        {rodriguez2015gremlin}
\bibfield{author}{\bibinfo{person}{Marko~A Rodriguez}.}
  \bibinfo{year}{2015}\natexlab{}.
\newblock \showarticletitle{The gremlin graph traversal machine and language
  (invited talk)}. In \bibinfo{booktitle}{\emph{Proceedings of the 15th
  Symposium on Database Programming Languages}}. \bibinfo{pages}{1--10}.
\newblock


\bibitem[Roy et~al\mbox{.}(2013)]%
        {roy2013x}
\bibfield{author}{\bibinfo{person}{Amitabha Roy}, \bibinfo{person}{Ivo
  Mihailovic}, {and} \bibinfo{person}{Willy Zwaenepoel}.}
  \bibinfo{year}{2013}\natexlab{}.
\newblock \showarticletitle{X-stream: Edge-centric graph processing using
  streaming partitions}. In \bibinfo{booktitle}{\emph{Proceedings of the
  Twenty-Fourth ACM Symposium on Operating Systems Principles}}.
  \bibinfo{pages}{472--488}.
\newblock


\bibitem[Shen et~al\mbox{.}(2023)]%
        {gart}
\bibfield{author}{\bibinfo{person}{Sijie Shen}, \bibinfo{person}{Zihang Yao},
  \bibinfo{person}{Lin Shi}, \bibinfo{person}{Lei Wang},
  \bibinfo{person}{Longbin Lai}, \bibinfo{person}{Qian Tao},
  \bibinfo{person}{Li Su}, \bibinfo{person}{Rong Chen},
  \bibinfo{person}{Wenyuan Yu}, \bibinfo{person}{Haibo Chen},
  \bibinfo{person}{Binyu Zang}, {and} \bibinfo{person}{Jingren Zhou}.}
  \bibinfo{year}{2023}\natexlab{}.
\newblock \showarticletitle{Bridging the Gap between Relational {OLTP} and
  Graph-based {OLAP}}. In \bibinfo{booktitle}{\emph{2023 USENIX Annual
  Technical Conference (USENIX ATC 23)}}. \bibinfo{publisher}{USENIX
  Association}, \bibinfo{pages}{181--196}.
\newblock


\bibitem[Su et~al\mbox{.}(2022)]%
        {su2022banyan}
\bibfield{author}{\bibinfo{person}{Li Su}, \bibinfo{person}{Xiaoming Qin},
  \bibinfo{person}{Zichao Zhang}, \bibinfo{person}{Rui Yang},
  \bibinfo{person}{Le Xu}, \bibinfo{person}{Indranil Gupta},
  \bibinfo{person}{Wenyuan Yu}, \bibinfo{person}{Kai Zeng}, {and}
  \bibinfo{person}{Jingren Zhou}.} \bibinfo{year}{2022}\natexlab{}.
\newblock \showarticletitle{Banyan: A Scoped Dataflow Engine for Graph Query
  Service}.
\newblock \bibinfo{journal}{\emph{Proc. VLDB Endow.}} \bibinfo{volume}{15},
  \bibinfo{number}{10} (\bibinfo{year}{2022}), \bibinfo{pages}{2045–2057}.
\newblock


\bibitem[Sun et~al\mbox{.}(2023)]%
        {Sun2023Legion}
\bibfield{author}{\bibinfo{person}{Jie Sun}, \bibinfo{person}{Li Su},
  \bibinfo{person}{Zuocheng Shi}, \bibinfo{person}{Wenting Shen},
  \bibinfo{person}{Zeke Wang}, \bibinfo{person}{Lei Wang}, \bibinfo{person}{Jie
  Zhang}, \bibinfo{person}{Yong Li}, \bibinfo{person}{Wenyuan Yu},
  \bibinfo{person}{Jingren Zhou}, {and} \bibinfo{person}{Fei Wu}.}
  \bibinfo{year}{2023}\natexlab{}.
\newblock \showarticletitle{Legion: Automatically Pushing the Envelope of
  {Multi-GPU} System for {Billion-Scale} {GNN} Training}. In
  \bibinfo{booktitle}{\emph{2023 USENIX Annual Technical Conference (USENIX ATC
  23)}}. \bibinfo{publisher}{USENIX Association}, \bibinfo{address}{Boston,
  MA}, \bibinfo{pages}{165--179}.
\newblock
\showISBNx{978-1-939133-35-9}


\bibitem[Sz\'{a}rnyas et~al\mbox{.}(2022)]%
        {snb-bi}
\bibfield{author}{\bibinfo{person}{G\'{a}bor Sz\'{a}rnyas},
  \bibinfo{person}{Jack Waudby}, \bibinfo{person}{Benjamin~A. Steer},
  \bibinfo{person}{D\'{a}vid Szak\'{a}llas}, \bibinfo{person}{Altan Birler},
  \bibinfo{person}{Mingxi Wu}, \bibinfo{person}{Yuchen Zhang}, {and}
  \bibinfo{person}{Peter Boncz}.} \bibinfo{year}{2022}\natexlab{}.
\newblock \showarticletitle{The LDBC Social Network Benchmark: Business
  Intelligence Workload}.
\newblock \bibinfo{journal}{\emph{Proc. VLDB Endow.}} \bibinfo{volume}{16},
  \bibinfo{number}{4} (\bibinfo{year}{2022}), \bibinfo{pages}{877–890}.
\newblock


\bibitem[Tian et~al\mbox{.}(2013)]%
        {tian2013think}
\bibfield{author}{\bibinfo{person}{Yuanyuan Tian}, \bibinfo{person}{Andrey
  Balmin}, \bibinfo{person}{Severin~Andreas Corsten}, \bibinfo{person}{Shirish
  Tatikonda}, {and} \bibinfo{person}{John McPherson}.}
  \bibinfo{year}{2013}\natexlab{}.
\newblock \showarticletitle{From" think like a vertex" to" think like a
  graph"}.
\newblock \bibinfo{journal}{\emph{Proceedings of the VLDB Endowment}}
  \bibinfo{volume}{7}, \bibinfo{number}{3} (\bibinfo{year}{2013}),
  \bibinfo{pages}{193--204}.
\newblock


\bibitem[Trigonakis et~al\mbox{.}(2021)]%
        {trigonakis2021adfs}
\bibfield{author}{\bibinfo{person}{Vasileios Trigonakis},
  \bibinfo{person}{Jean-Pierre Lozi}, \bibinfo{person}{Tom{\'a}{\v{s}}
  Falt{\'\i}n}, \bibinfo{person}{Nicholas~P Roth}, \bibinfo{person}{Iraklis
  Psaroudakis}, \bibinfo{person}{Arnaud Delamare}, \bibinfo{person}{Vlad
  Haprian}, \bibinfo{person}{C{\u{a}}lin Iorgulescu}, \bibinfo{person}{Petr
  Koupy}, \bibinfo{person}{Jinsoo Lee}, {et~al\mbox{.}}}
  \bibinfo{year}{2021}\natexlab{}.
\newblock \showarticletitle{$\{$aDFS$\}$: An Almost $\{$Depth-First-Search$\}$
  Distributed $\{$Graph-Querying$\}$ System}. In \bibinfo{booktitle}{\emph{2021
  USENIX Annual Technical Conference (USENIX ATC 21)}}.
  \bibinfo{pages}{209--224}.
\newblock


\bibitem[Tu et~al\mbox{.}(2023)]%
        {tu2023disentangled}
\bibfield{author}{\bibinfo{person}{Ke Tu}, \bibinfo{person}{Wei Qu},
  \bibinfo{person}{Zhengwei Wu}, \bibinfo{person}{Zhiqiang Zhang},
  \bibinfo{person}{Zhongyi Liu}, \bibinfo{person}{Yiming Zhao},
  \bibinfo{person}{Le Wu}, \bibinfo{person}{Jun Zhou}, {and}
  \bibinfo{person}{Guannan Zhang}.} \bibinfo{year}{2023}\natexlab{}.
\newblock \showarticletitle{Disentangled Interest importance aware Knowledge
  Graph Neural Network for Fund Recommendation}. In
  \bibinfo{booktitle}{\emph{Proceedings of the 32nd ACM International
  Conference on Information and Knowledge Management}}.
  \bibinfo{pages}{2482--2491}.
\newblock


\bibitem[Vora(2019)]%
        {vora2019lumos}
\bibfield{author}{\bibinfo{person}{Keval Vora}.}
  \bibinfo{year}{2019}\natexlab{}.
\newblock \showarticletitle{$\{$LUMOS$\}$:$\{$Dependency-Driven$\}$ Disk-based
  Graph Processing}. In \bibinfo{booktitle}{\emph{2019 USENIX Annual Technical
  Conference (USENIX ATC 19)}}. \bibinfo{pages}{429--442}.
\newblock


\bibitem[Wang et~al\mbox{.}(2023)]%
        {wang2023neural}
\bibfield{author}{\bibinfo{person}{Xiyuan Wang}, \bibinfo{person}{Haotong
  Yang}, {and} \bibinfo{person}{Muhan Zhang}.} \bibinfo{year}{2023}\natexlab{}.
\newblock \bibinfo{title}{Neural Common Neighbor with Completion for Link
  Prediction}.
\newblock
\newblock
\showeprint[arxiv]{2302.00890}~[cs.LG]


\bibitem[Wang et~al\mbox{.}(2017)]%
        {wang2017gunrock}
\bibfield{author}{\bibinfo{person}{Yangzihao Wang}, \bibinfo{person}{Yuechao
  Pan}, \bibinfo{person}{Andrew Davidson}, \bibinfo{person}{Yuduo Wu},
  \bibinfo{person}{Carl Yang}, \bibinfo{person}{Leyuan Wang},
  \bibinfo{person}{Muhammad Osama}, \bibinfo{person}{Chenshan Yuan},
  \bibinfo{person}{Weitang Liu}, \bibinfo{person}{Andy~T Riffel},
  {et~al\mbox{.}}} \bibinfo{year}{2017}\natexlab{}.
\newblock \showarticletitle{Gunrock: GPU graph analytics}.
\newblock \bibinfo{journal}{\emph{ACM Transactions on Parallel Computing
  (TOPC)}} \bibinfo{volume}{4}, \bibinfo{number}{1} (\bibinfo{year}{2017}),
  \bibinfo{pages}{3}.
\newblock


\bibitem[Yang et~al\mbox{.}(2022)]%
        {yang2022gnnlab}
\bibfield{author}{\bibinfo{person}{Jianbang Yang}, \bibinfo{person}{Dahai
  Tang}, \bibinfo{person}{Xiaoniu Song}, \bibinfo{person}{Lei Wang},
  \bibinfo{person}{Qiang Yin}, \bibinfo{person}{Rong Chen},
  \bibinfo{person}{Wenyuan Yu}, {and} \bibinfo{person}{Jingren Zhou}.}
  \bibinfo{year}{2022}\natexlab{}.
\newblock \showarticletitle{GNNLab: a factored system for sample-based GNN
  training over GPUs}. In \bibinfo{booktitle}{\emph{Proceedings of the
  Seventeenth European Conference on Computer Systems}}.
  \bibinfo{pages}{417--434}.
\newblock


\bibitem[Yang et~al\mbox{.}(2021)]%
        {yang2021consisrec}
\bibfield{author}{\bibinfo{person}{Liangwei Yang}, \bibinfo{person}{Zhiwei
  Liu}, \bibinfo{person}{Yingtong Dou}, \bibinfo{person}{Jing Ma}, {and}
  \bibinfo{person}{Philip~S Yu}.} \bibinfo{year}{2021}\natexlab{}.
\newblock \showarticletitle{Consisrec: Enhancing gnn for social recommendation
  via consistent neighbor aggregation}. In
  \bibinfo{booktitle}{\emph{Proceedings of the 44th international ACM SIGIR
  conference on Research and development in information retrieval}}.
  \bibinfo{pages}{2141--2145}.
\newblock


\bibitem[Yi et~al\mbox{.}(2022)]%
        {yi2022graph}
\bibfield{author}{\bibinfo{person}{Hai-Cheng Yi}, \bibinfo{person}{Zhu-Hong
  You}, \bibinfo{person}{De-Shuang Huang}, {and} \bibinfo{person}{Chee~Keong
  Kwoh}.} \bibinfo{year}{2022}\natexlab{}.
\newblock \showarticletitle{Graph representation learning in bioinformatics:
  trends, methods and applications}.
\newblock \bibinfo{journal}{\emph{Briefings in Bioinformatics}}
  \bibinfo{volume}{23}, \bibinfo{number}{1} (\bibinfo{year}{2022}),
  \bibinfo{pages}{bbab340}.
\newblock


\bibitem[Yu et~al\mbox{.}(2023)]%
        {vineyard}
\bibfield{author}{\bibinfo{person}{Wenyuan Yu}, \bibinfo{person}{Tao He},
  \bibinfo{person}{Lei Wang}, \bibinfo{person}{Ke Meng}, \bibinfo{person}{Ye
  Cao}, \bibinfo{person}{Diwen Zhu}, \bibinfo{person}{Sanhong Li}, {and}
  \bibinfo{person}{Jingren Zhou}.} \bibinfo{year}{2023}\natexlab{}.
\newblock \showarticletitle{Vineyard: Optimizing Data Sharing in Data-Intensive
  Analytics}.
\newblock \bibinfo{journal}{\emph{Proc. ACM Manag. Data}} \bibinfo{volume}{1},
  \bibinfo{number}{2} (\bibinfo{year}{2023}).
\newblock


\bibitem[Zamini et~al\mbox{.}(2022)]%
        {zamini2022review}
\bibfield{author}{\bibinfo{person}{Mohamad Zamini}, \bibinfo{person}{Hassan
  Reza}, {and} \bibinfo{person}{Minou Rabiei}.}
  \bibinfo{year}{2022}\natexlab{}.
\newblock \showarticletitle{A review of knowledge graph completion}.
\newblock \bibinfo{journal}{\emph{Information}} \bibinfo{volume}{13},
  \bibinfo{number}{8} (\bibinfo{year}{2022}), \bibinfo{pages}{396}.
\newblock


\bibitem[Zhang and Chartrand(2006)]%
        {zhang2006introduction}
\bibfield{author}{\bibinfo{person}{P Zhang} {and} \bibinfo{person}{G
  Chartrand}.} \bibinfo{year}{2006}\natexlab{}.
\newblock \bibinfo{booktitle}{\emph{Introduction to graph theory}}.
\newblock \bibinfo{publisher}{Tata McGraw-Hill}.
\newblock


\bibitem[Zhang et~al\mbox{.}(2021)]%
        {zhang2021graph}
\bibfield{author}{\bibinfo{person}{Xiao-Meng Zhang}, \bibinfo{person}{Li
  Liang}, \bibinfo{person}{Lin Liu}, {and} \bibinfo{person}{Ming-Jing Tang}.}
  \bibinfo{year}{2021}\natexlab{}.
\newblock \showarticletitle{Graph neural networks and their current
  applications in bioinformatics}.
\newblock \bibinfo{journal}{\emph{Frontiers in genetics}}  \bibinfo{volume}{12}
  (\bibinfo{year}{2021}), \bibinfo{pages}{690049}.
\newblock


\bibitem[Zhou et~al\mbox{.}(2020)]%
        {zhou2020graph}
\bibfield{author}{\bibinfo{person}{Jie Zhou}, \bibinfo{person}{Ganqu Cui},
  \bibinfo{person}{Shengding Hu}, \bibinfo{person}{Zhengyan Zhang},
  \bibinfo{person}{Cheng Yang}, \bibinfo{person}{Zhiyuan Liu},
  \bibinfo{person}{Lifeng Wang}, \bibinfo{person}{Changcheng Li}, {and}
  \bibinfo{person}{Maosong Sun}.} \bibinfo{year}{2020}\natexlab{}.
\newblock \showarticletitle{Graph neural networks: A review of methods and
  applications}.
\newblock \bibinfo{journal}{\emph{AI open}}  \bibinfo{volume}{1}
  (\bibinfo{year}{2020}), \bibinfo{pages}{57--81}.
\newblock


\bibitem[Zhou et~al\mbox{.}(2023)]%
        {hierarchical-knowledge-graph}
\bibfield{author}{\bibinfo{person}{Zhilun Zhou}, \bibinfo{person}{Yu Liu},
  \bibinfo{person}{Jingtao Ding}, \bibinfo{person}{Depeng Jin}, {and}
  \bibinfo{person}{Yong Li}.} \bibinfo{year}{2023}\natexlab{}.
\newblock \showarticletitle{Hierarchical Knowledge Graph Learning Enabled
  Socioeconomic Indicator Prediction in Location-Based Social Network}
  \emph{(\bibinfo{series}{WWW '23})}. \bibinfo{publisher}{Association for
  Computing Machinery}, \bibinfo{address}{New York, NY, USA},
  \bibinfo{pages}{122–132}.
\newblock
\showISBNx{9781450394161}


\bibitem[Zhu et~al\mbox{.}(2016)]%
        {zhu2016gemini}
\bibfield{author}{\bibinfo{person}{Xiaowei Zhu}, \bibinfo{person}{Wenguang
  Chen}, \bibinfo{person}{Weimin Zheng}, {and} \bibinfo{person}{Xiaosong Ma}.}
  \bibinfo{year}{2016}\natexlab{}.
\newblock \showarticletitle{Gemini: A Computation-Centric Distributed Graph
  Processing System.}. In \bibinfo{booktitle}{\emph{OSDI}}.
  \bibinfo{pages}{301--316}.
\newblock


\bibitem[Zhu et~al\mbox{.}(2019)]%
        {zhu13livegraph}
\bibfield{author}{\bibinfo{person}{Xiaowei Zhu}, \bibinfo{person}{Guanyu Feng},
  \bibinfo{person}{Marco Serafini}, \bibinfo{person}{Xiaosong Ma},
  \bibinfo{person}{Jiping Yu}, \bibinfo{person}{Lei Xie},
  \bibinfo{person}{Ashraf Aboulnaga}, {and} \bibinfo{person}{Wenguang Chen}.}
  \bibinfo{year}{2019}\natexlab{}.
\newblock \showarticletitle{LiveGraph: A Transactional Graph Storage System
  with Purely Sequential Adjacency List Scans}.
\newblock \bibinfo{journal}{\emph{Proceedings of the VLDB Endowment}}
  \bibinfo{volume}{13}, \bibinfo{number}{7} (\bibinfo{year}{2019}).
\newblock


\end{thebibliography}

\end{document}